\documentclass[12pt,preprint]{aastex}

\def\eg{\mbox{e.g.}}
\def\ie{\mbox{i.e.}}
\def\Msun{\mbox{$M_{\odot}$}}

\usepackage{natbib}
\usepackage{graphicx}

\slugcomment{Accepted for publication in ApJS}

\shorttitle{Radial Velocities and Pulsation Ephemerides of 11 Field RR Lyrae Stars}
\shortauthors{B.-Q. For, G. W. Preston and C. Sneden}

\begin{document}

\title{Radial Velocities and Pulsation Ephemerides of 11 Field RR Lyrae Stars}

\author{Bi-Qing For\altaffilmark{1}, George W. Preston\altaffilmark{2}, Christopher Sneden\altaffilmark{1}}

\altaffiltext{1}{Department of Astronomy, University of Texas, Austin, TX 78712, USA; biqing@astro.as.utexas.edu}
\altaffiltext{2}{Carnegie Observatories, 813 Santa Barbara Street, Pasadena, CA 91101, USA}

\begin{abstract}

We present new radial velocities, improved pulsation periods and reference epochs of 11 field 
RR Lyrae ab-type variables: AS Vir, BS Aps, CD Vel, DT Hya,  
RV Oct, TY Gru, UV Oct, V1645 Sgr, WY Ant, XZ Aps and Z Mic. 
This study is based on high resolution spectra obtained with the echelle spectrograph 
of the 2.5-m du Pont telescope at Las Campanas Observatory. We obtained $\sim 200$ spectra 
per star (\ie, total of $\sim 2300$ spectra) distributed more or 
less uniformly throughout their pulsation cycles.
Radial velocity curves and photometric lightcurves phased to our new ephemerides are presented 
for all program stars. In a subsequent paper, 
we will use these spectra to derive stellar atmospheric parameters and chemical compositions 
throughout the pulsational cycles, based purely on spectroscopic constraints. 
\end{abstract}

\keywords{stars: horizontal-branch -- stars: Population II -- stars: variables: RR Lyrae -- techniques: radial velocities}

\section{Introduction}

RR Lyraes (RR Lyr), named after their prototype, 
are old, low-mass stars that reside in the instability strip of 
the horizontal branch (HB). 
They are powerful tools in the studies of many fundamental astrophysical problems. 
Due to their variability and relatively high luminosity, they are easily 
identified even at large distances. Their small dispersion in intrinsic mean luminosity 
makes them good standard candles in contrast to other stellar tracers, 
such as M giants \citep{Majewski03}. In addition to the distance scale, 
RR Lyr play an important role in 
studying Galactic structure and formation. 
They are generally used to trace the spatial and kinematic distribution of 
the old stellar populations of the Galactic disk and halo components. For example, 
recent optical RR Lyrae surveys, 
such as QUEST \citep{Vivas04} and SDSS \citep{Ivezic04}, 
have revealed halo substructures and dynamically young stellar streams 
that are associated with the formation of the outer halo.  

RR Lyr are also commonly used to study the chemical evolution of the disk and halo of 
our Milky Way. This effort began with the pioneering low-resolution spectroscopic survey 
by \citet{Preston59}, who introduced the $\Delta S$ index that describes the relation 
between Hydrogen and calcium $K$-line absorption strengths. The $\Delta S$ index varies 
during RR Lyrs pulsational cycle, so the standard $\Delta S$ index is defined 
at light minimum (\ie, near phase 0.8). High-resolution studies 
generally have concentrated on limited 
phases near minimum light, because of the relatively slow variations in photometric 
effective temperature that occur at these pulsation phases. 

Our work in this area began as an investigation of the systematics of chemical 
abundances along the HB in the Galactic Halo \citep{FS10}. 
The primary objectives of that paper were to investigate any abundance anormalies 
in non-variable RHB and BHB stars, to derive masses of these stars and to 
determine the red and blue edges of the RR Lyrae instability strip. 
They concluded that: (1) the abundance ratios of these stars are generally consistent 
with those of similar-metallicity field stars in different evolutionary stages, 
(2) the stars possess masses of $\sim0.5$ \Msun, and (3) the 
effective temperatures for the red and blue edges of HB stars in the metallicity 
range $-0.8 \gtrsim$ [Fe/H] $\gtrsim\-2.5$ are 5900 K and 7400 K, respectively. 

We are applying the analytical techniques of \citet{FS10} to a controlled 
sample of RR Lyr stars. The spectra have been gathered by GWP 
for his investigation of many issues in RR Lyr atmospheric dynamics, such as shocks, 
turbulent and Blazhko effect. 
This RR Lyr spectral study was also initiated partly to better understand 
the nature of a carbon-rich and $s$-process rich RRab star, 
TY Gru \citep{Preston06}. This star was identified as CS~22881$-$071 in 
the HK objective-prism survey \citep{BPS92} and was 
initially included in the study of chemical abundance of a sample metal-poor 
red horizontal branch stars \citep{PrestonRHB06}. 
The enrichment of carbon and $n$-capture species suggests that this star might 
have gone through binary mass transfer from a primary star during its Asymptotic Giant 
Branch (AGB) evolution (\citealp{Preston06} and references therein). 
To further investigate the abundance anomalies as seen in TY Gru 
and to detect the possible orbital motion caused by the relic companion of an 
AGB star \citep{Preston11}, GWP selected 
a sample of RRab stars with $P\sim 0.57$ day that are broadly representative of the metal-poor halo. 
Numerous observations at all pulsation phases provide a dataset that can be used 
to investigate the dependence of derived abundances on the various 
thermodynamic conditions that occur during pulsation cycles. 

In this paper, we present radial velocities (RVs) and improved ephemerides 
of 11 field RRab stars. In a subsequent paper, we will report  
stellar parameter and chemical abundance analyses throughout the 
pulsational cycles (For, Sneden $\&$ Preston, in preparation). We provide the 
basic information on targets and describe the observations and reduction 
methods in \S 2 and \S 3. In \S 4, we present the derived radial velocities 
and improved ephemerides. 

\section{Targets and Observations}

The observations were made with echelle spectrograph of the du Pont 2.5-m telescope at the 
Las Campanas Observatory (LCO) during 2006--2009.
We used this instrument configured with the $1.5\arcsec\times4\arcsec$ entrance slit, 
which gives a resolving power of $R \equiv \lambda/\Delta\lambda \sim 27,000$ 
at the Mg I b lines (5180 \AA). 
The total wavelength coverage is $3500-9000$ \AA. Integration times ranged from 
a minimum value of 200~s (to insure reasonably uniform illumination of the slit 
by starlight) to an upper limit of 600~s (to avoid excessive blurring of the spectrum 
due to changing radial velocity). 
The values of S/N achieved by such integrations can be estimated by 
observations of CS~22175$-$034 \citep{Preston91}, which is a star with similar colors to RR Lyrs. 
Spectra of this star ($V=12.60$, $B-V=0.37$) obtained near the zenith under typical observing conditions 
with an exposure time of 600~s achieved S/N$\sim10$ at 4050\AA, 
S/N$\sim15$ at 4300\AA, 
S/N$\sim20$ at 5000\AA, 
S/N$\sim30$ at 6000\AA\,and 
S/N$\sim30$ at 6600\AA.   
Wavelength calibrations were achieved by taking Thorium-Argon comparison lamp exposures at least once 
per hour at each star position. 
Basic information about our program stars is given in Table~\ref{targets}.

\section{Data Reduction}

The raw data were bias subtracted, flat-fielded, background subtracted, 
then extracted to one-dimensional (1D) spectra and wavelength-calibrated by use of 
IRAF\footnote{The Image Reduction and Analysis Facility, a general purpose 
software package for astronomical data, written and supported by the 
IRAF programming group of the National Optical Astronomy Observatories 
(NOAO) in Tucson, AZ.} ECHELLE package. Thorium-Argon identifications 
were based on the line list in the IRAF package data file 
(thar.dat) and the Th-Ar wavelength table\footnote{http://www.lco.cl/telescopes-information/irenee-du-pont/instruments/website/echelle-spectrograph-manuals/echelle-spectrograph-manuals/atlas} provided by the LCO. 
We paid particular attention to scattered light corrections, and 
in the following subsection we describe our own (non-global) approach to this problem. 

\subsection{Scattered Light Correction}

Some of the incident photons at each wavelength are scattered into 
all echelle orders by optical imperfections 
in the optical train of the spectrograph.  
A generic method for making scattered light
corrections is use of the IRAF apscatter task, in which the scattered light 
pixels are fitted by a series of 1D functions across the dispersion. 
The independent fits are then smoothed along the dispersion by again fitting low order functions. 
These fits then define the smooth scattered light surface to be subtracted from the image.  
Application of this method to du Pont echelle spectra is complicated by a number 
of considerations discussed below.

A fraction of the photons of every wavelength that passed through the $1.5\arcsec\times4.0\arcsec$ 
entrance slit were scattered into the image plane of the du Pont spectrograph by imperfect 
transmission/reflection at surfaces in the optical train. 
Longward of 6500 \AA\,the inter-order space became too small to measure pure scattered light. 
To circumvent this difficulty, we obtained observations of 4 standard stars through a 
small $0.75\arcsec\times0.75\arcsec$, for which the inter-order space was more than adequate. 
Additional difficulties in data reduction arised due to our adopted observing procedure.  
Long experience at the du Pont had shown that accurate sky subtraction could not be achieved by use 
of light adjacent to the star image because of centering and guiding errors. 
If sky background is important, it must be measured by sky observations before and/or after the stellar observation, 
and only under good photometric conditions.  
For stars brighter than magnitude 13, sky was unimportant at the 1$\%$ level except near the full moon, 
which we avoided, so we ignored it.  
To save the precious time between observations that would be required to rotate the spectrograph, 
we made all observations with an east-west oriented slit. 
Furthermore, we guided with a red-sensitive detector, 
so that at many telescope positions significant fractions of blue-violet light did not pass through the slit 
due to atmospheric dispersion: 
the observed spectra were thus somewhat "reddened" and mimicked those of lower color temperature. 
In addition, this could affect the velocity difference between the red and blue lines. 

To investigate such affect on our spectral line width, 
we calculated the velocity shifts between spectral regions 
at 4000--6000~\AA\ using the followings: 
(1) calculate the parallactic angle for each of our stars 
at different 7 hour angles (from 0.01--6~hr with increment of 1~hr); 
(2) calculate the angle between east--west slit of Cassegrain 
spectrograph on equatorially--mounted telescope and direction to zenith; 
(3) calculate the sine of the inclination of the spectrum to the slit;
(4) calculate the altitude of each star;
(5) calculate differential atmospheric dispersion by linear approximate 
of data shown in Figure~2 of \citet{Simon66} for elevation 2811~m, which 
is close to du Pont elevation at 2200~m; 
(6) calculate the differential atmospheric dispersion perpendicular 
to the slit;
(7) finally, convert angular displacement in arcsec to velocity displacement 
by use of scale factor 8 km~s$^{-1}$~arcsec$^{-1}$ (assuming 1.5$\arcsec$ 
slit width projects to 12 km~s$^{-1}$) The upper limit of 6 km~s$^{-1}$ 
was set for the conversion, which corresponds to an illumination centroid 
at the edge of the slit. During the observation, seeing and guiding 
errors will diminish atmospheric displacements, \eg, 
producing centroids nearer to aperture center. 
Our velocity displacement calculations over  7 hr angles of each star 
range from 0--2 km~s$^{-1}$, which are small compared to intrinsic 
RR Lyr line width of $> 20$ km~s$^{-1}$. 
Thus, the broadening effects of these displacements are small on 
individual spectra. 

To further investigate if such broadening would have any affect on the co-added spectrum, 
we measured the equivalent widths of several metal lines 
of individual spectrum and co-added spectrum of the same phase. 
The comparisons of measured equivalent width are consistent with an overall 
difference of $\pm3$ m\AA. As such, we conclude that the equivalent widths 
are unaffected in these cases. However, we warn the reader that 
the shifts certainly contribute to systematic errors of individual 
radial velocities, especially for stars with large Sourthern declinations 
(DEC $< -70$). Inspecting the scatter of RV data for the stable RRab stars 
(WY Ant, DT Hya, CD Vel, XZ Aps, RV Oct and Zmic) of our RV curves, 
the errors due to blue image decentering cannot be much greater than 
1 km~s$^{-1}$.

The raw spectra of the observed standard stars were bias-subtracted and flat fielded. 
Then, individual spectra were combined into a single spectrum. 
We extracted each combined spectrum with 6 pixel aperture 
to two 1D spectra with one for star and one for inter-order background.
Subsequently, the 1D spectra were continuum normalized with the continuum task in IRAF ECHELLE package. 
To obtain the contribution of scattered light in each order, we calculated the 
fractional contribution of the inter-order background light to the on-order starlight 
as a function of spectral order (or wavelength), b$_{\lambda}$/s$_{\lambda}$, for each standard star. 
Because a 10 pixel aperture was used to extract the spectra of our program RR Lyrs, 
the extracted 1D spectra are expected to contain 
more scattered light than the extracted scattered light frames. 
Thus, we applied a correction factor of 5/3 to the calculated b$_{\lambda}$/s$_{\lambda}$ ratios.

\begin{figure}
\begin{center}
\includegraphics[scale=0.5,angle=-90]{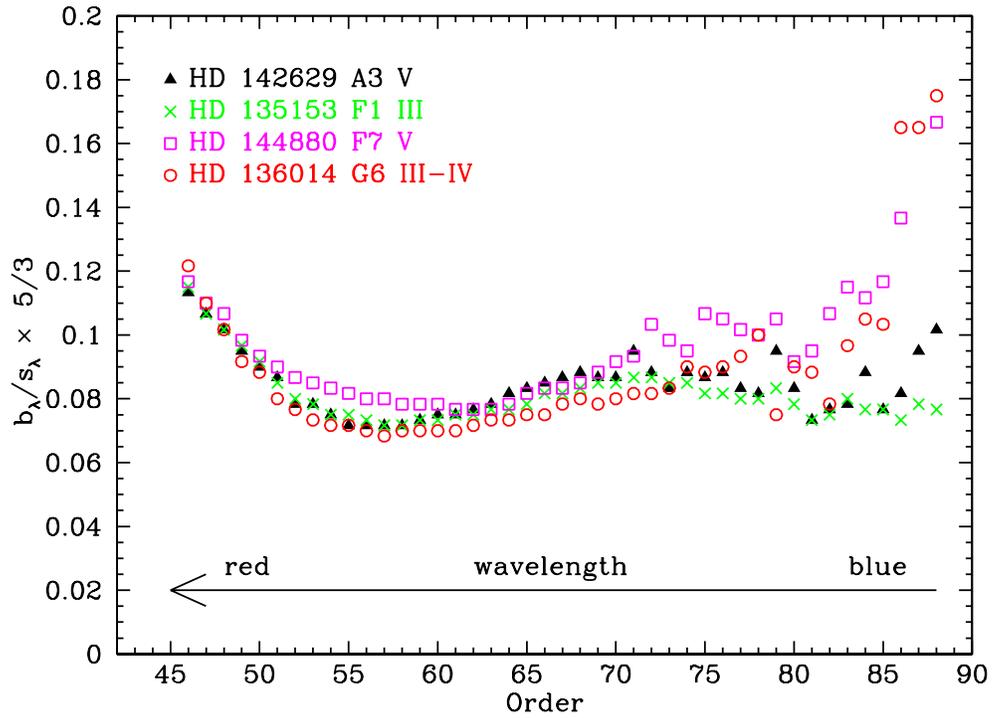}
\caption{Fractional contribution of the inter-order background light to the on-order starlight 
as a function of spectral order (wavelength), b$_{\lambda}$/s$_{\lambda}$, for each standard star. 
Wavelength decreases with increasing order.  \label{ratios_f}}
\end{center}
\end{figure}

In Table~\ref{standard}, we provide the basic information and observing log for our standard stars. 
The calculated fraction as a function of spectral order for each standard star is presented 
in Table~\ref{ratios_t}. 
We summarize our results in Figure~\ref{ratios_f}. 
The success of this calibration procedure depends on the stability of the scattered light distribution 
produced by the spectrograph. Recalibration performed from time to time by the procedure described 
above has shown that the scattered light distribution has changed little, if at all, during the past two decades. 
We will conisder this issue more in the subsequent paper. 

\section{Analysis}

\subsection{Radial Velocities}

The spectra that we used for deriving the RVs were not corrected for scattered light. 
It is not important for deriving the RVs but will be required for the the subsequent 
atmospheric analysis. 
We derived the RVs by use of the cross-correlation FXCOR task in IRAF, in which 
the individual spectra were cross-correlated against 
a template by fitting a Gaussian to the cross-correlation peak. We 
constructed the individual spectra from 13 echelle orders covering 
the spectral region of $4000-4600$ \AA, which were then flattened, normalized, and 
stitched together with an IRAF script.     
In order to get strong cross-correlations that minimizes RV errors, we 
created a template from several spectra of CS~22874$-$009, 
a blue metal-poor radial velocity standard star \citep{PS00}, which possesses a spectrum similar 
to those of RR Lyr stars at most phases. 
The typical RV error calculated from FXCOR is $\sim0.5$ km~s$^{-1}$.
We present the observed HJD midpoints, 
phases (see \S 4.2), derived RVs and their associated errors in Table~\ref{RVs}. 

\subsection{Pulsation Ephemerides}

A pulsation ephemeris is commonly written as HJD~(max~light) = $T_{0}$ + $n\times$P,
where  $T_{0}$ is epoch, $n$ is the number of elapsed 
pulsation cycles and $P$ is the pulsational period in days. 
The All Sky Automated Survey\footnote{http://www.astrouw.edu.pl/asas/} 
(ASAS) \citep{Pojmanski02} provides 
a starting point to obtain ephemerides for our program stars. 
This photometric survey has been carried out over many years 
at the LCO and Haleakala, Maui stations. 
Using the ASAS reported pulsation period 
and $T_{0}$, the folded lightcurves as shown on the ASAS website were slightly out of phase. 
This suggests that the quoted values could be improved.  
Here we present the methods of improving both pulsation periods and 
$T_{0}$ values of our program stars. 

\subsubsection{Pulsation Period}

We improved the pulsation periods of our 10 RRab stars using the 
classified ``grade A'' $V$-band photometric data listed in the ASAS database. 
The pulsation period of TY Gru was adopted from \citet{Preston06} since those authors 
derived it by use of additional observations obtained with the LCO Swope telescope. 
The pulsational periods were derived using the Lomb-Scargle periodogram \citep{Scargle82}. 
We set a short period range of 0.5--0.6 day to minimize the chance of selecting spurious peaks 
caused by aliasing sidelobes (due to large observational gap and unevenly spaced time series data) 
in a different frequency domain. 
In addition, the pulsational period of our RRab stars is known within this range, 
in which a smaller time step can be set to achieve accuracy while cutting down the computing time. 
The advantages of this algorithm are: 
(1) less computing time than the lightcurve template fitting method, which requires 
continuous sampled data sets that are not available from the ASAS database and 
(2) the ability to compute Fourier Transform for unevenly spaced time series data. 
While we have continuous sampled RV data, we still cannot use 
the template fitting method because it is designed for lightcurves fitting not for 
RV curves of RR Lyr stars. We warn the reader that there is a caveat for this algorithm. 
It is optimized to identify sinusoidal-shaped periodic signal in time-series data. 
The lightcurves of RR Lyrae stars are non-sinusoidal. 

\begin{figure}
\begin{center}
\includegraphics[scale=0.5,angle=-90]{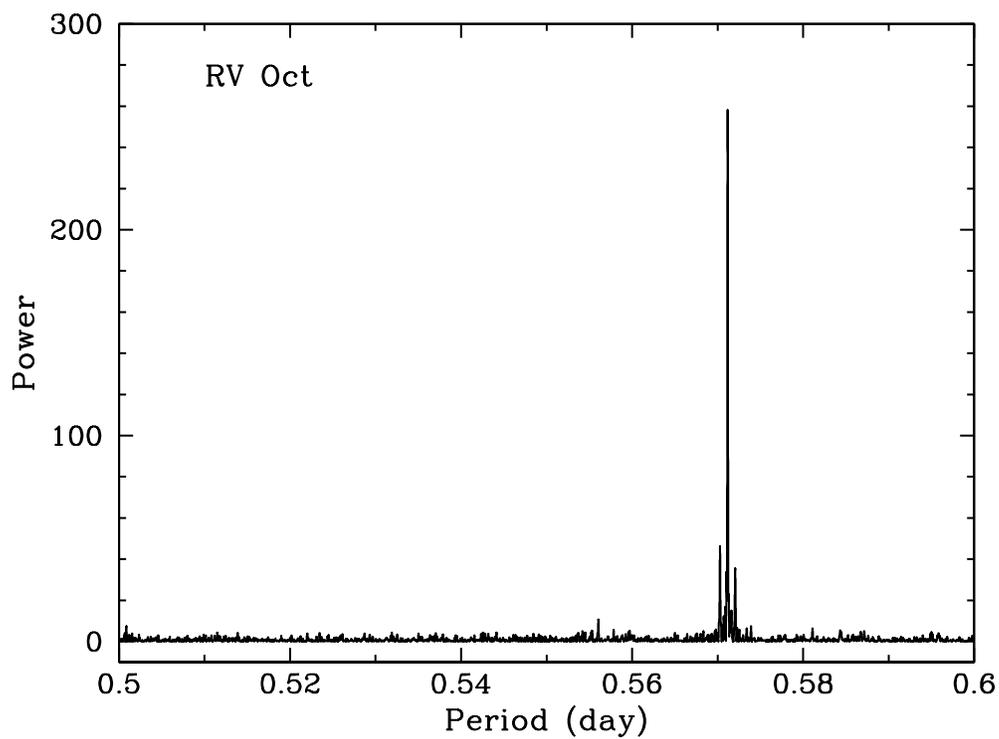}
\caption{An example of typical periodogram used for searching the pulsational period. 
The sidelobes that caused by the large observing gap is clearly seen in the periodogram. 
The highest peak defines the pulsational period of RV Oct. \label{periodogram}}
\end{center}
\end{figure}

Given that we have huge amount of photometric data, the highest peak, 
which represents the most probable repeating signal, in a periodogram is 
always more than $4\sigma$ above the mean noise level (see Figure~\ref{periodogram}). 
The highest peak of each periodogram is selected as the pulsational period of 
our program stars. We evaluated the error of the periods by comparing the periods 
derived from Lomb-Scargle algorithm and Box-fitting least squares method (BLS) \citep{Kovacs02}. 
The BLS algorithm fits the input time series with ``box''-shaped function, which 
makes it more suitable for obtaining period for transiting lightcurve 
than RR Lyrs lightcurve. 

In Table~\ref{eph}, we present the pulsation periods quoted 
in the ASAS catalog in column 5, the derived pulsation periods and their 
associated errors in column 6 and 7. The error of the period is within 
0.000001--0.000007 day, which is 10 times better than the periods accuracy 
quoted at the ASAS website.

\subsubsection{Epoch}

The reference epoch ($T_{0}$) of a pulsating variable star is 
usually chosen to occur at visual light maximum, 
which closely coincides with RV minimum (see discussion by \citealp{Preston09} and particularly 
Figures~\ref{RV-CDVel}--\ref{RV-TYGru}). 
Because the periodogram does not calculate an epoch, we derived values of $T_{0}$ by 
use of the Kwee-van Woerden method \citep{KVW56}. 
This method is generally used for computing the epoch of minimum of 
eclipsing variables accurately but it is also suitable to determine the epoch of 
light maxima of variable stars. 
We prefer to use our RV curves for this purpose because adequate data points 
near the RV minima (light maxima) were available during individual cycles, 
in constrast to the ASAS lightcurve data that were collected over long time 
intervals, with few observations per cycle and relatively 
large scatter near light maxima. 

For each star, we selected the cycles that cover the RV minima and calculated 
several equidistant midpoints between the rising and descending branch near the RV 
minima for a given cycle. Then, we fitted 
a linear least square equation to these midpoints, which the intersection 
of the straight line and the RV curve gives the $T_{0}$ of RV minima. 
We typically computed more than one $T_{0}$ using the above method 
per star to evaluate the error. Assuming the pulsational period and 
the first derived $T_{0}$ are accurate, we can calculate the predicted 
$T_{0}$ after $n$ pulsation cycles using the defined pulsation ephemeris above. The predicted $T_{0}$ 
should be close to the second derived $T_{0}$. The difference between 
the predicted and the derived value provides an estimate for the error. 
Due to the possibility of period change for the Blazhko RRab stars, 
several epochs were determined and used for folding their RV curves. 

\begin{figure}
\begin{center}
\includegraphics[scale=0.5,angle=-90]{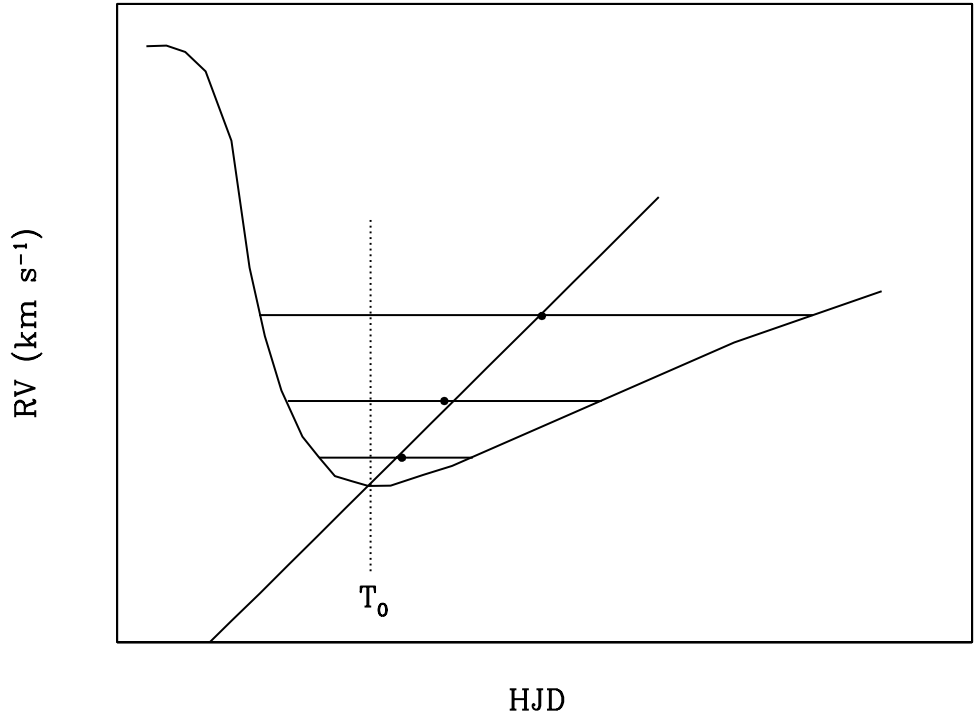}
\caption{A schematic diagram that shows the typical radial velocity curve near minima (or equivalent to 
light maxima) of a RRab variable star. It also shows the Kwee-van Woerden method \citep{KVW56} that 
we applied to determine the epochs of our RR Lyrae stars. \label{kvwdiagram}}
\end{center}
\end{figure}

\begin{figure}
\begin{center}
\includegraphics[scale=0.45,angle=-90]{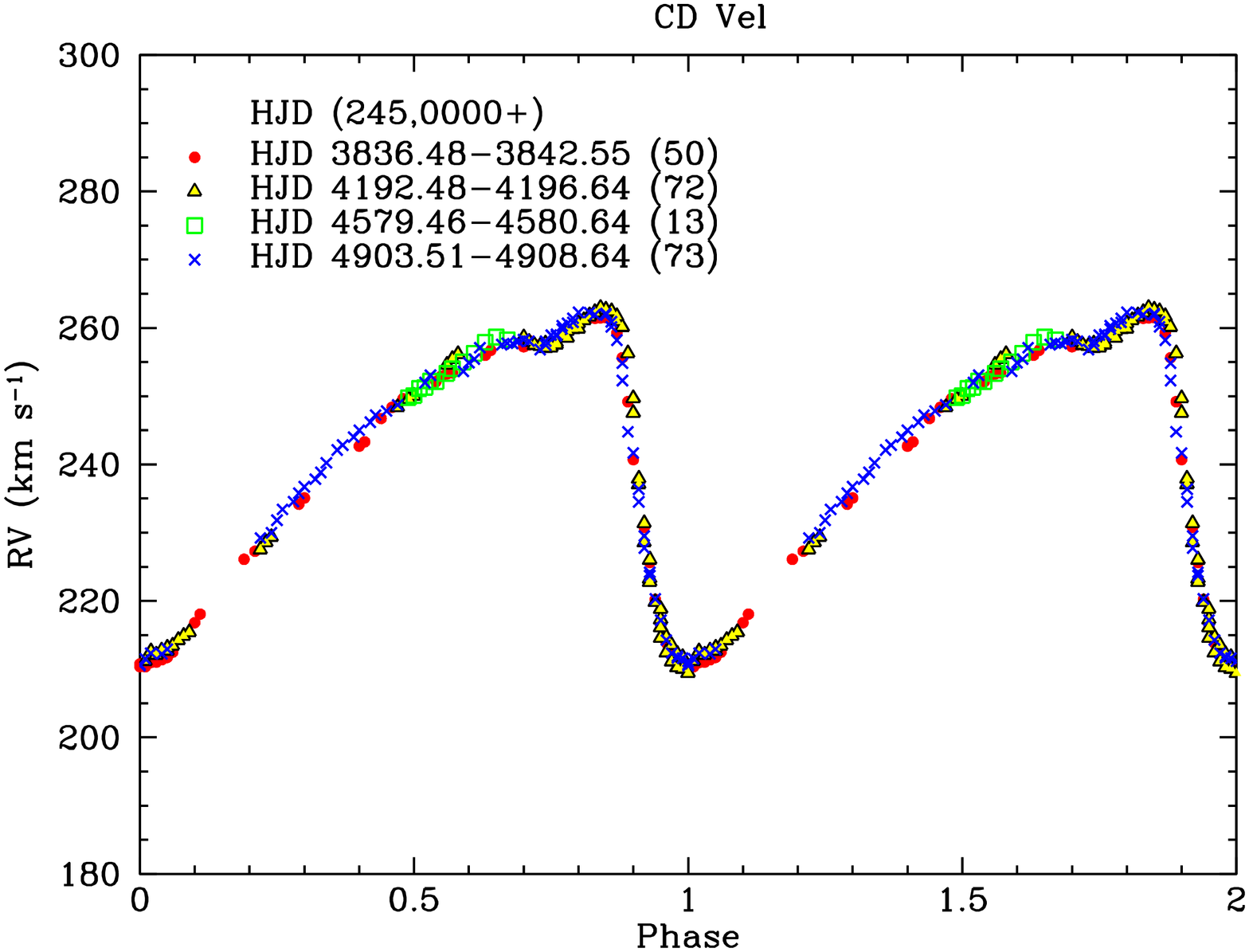}
\includegraphics[scale=0.45,angle=-90]{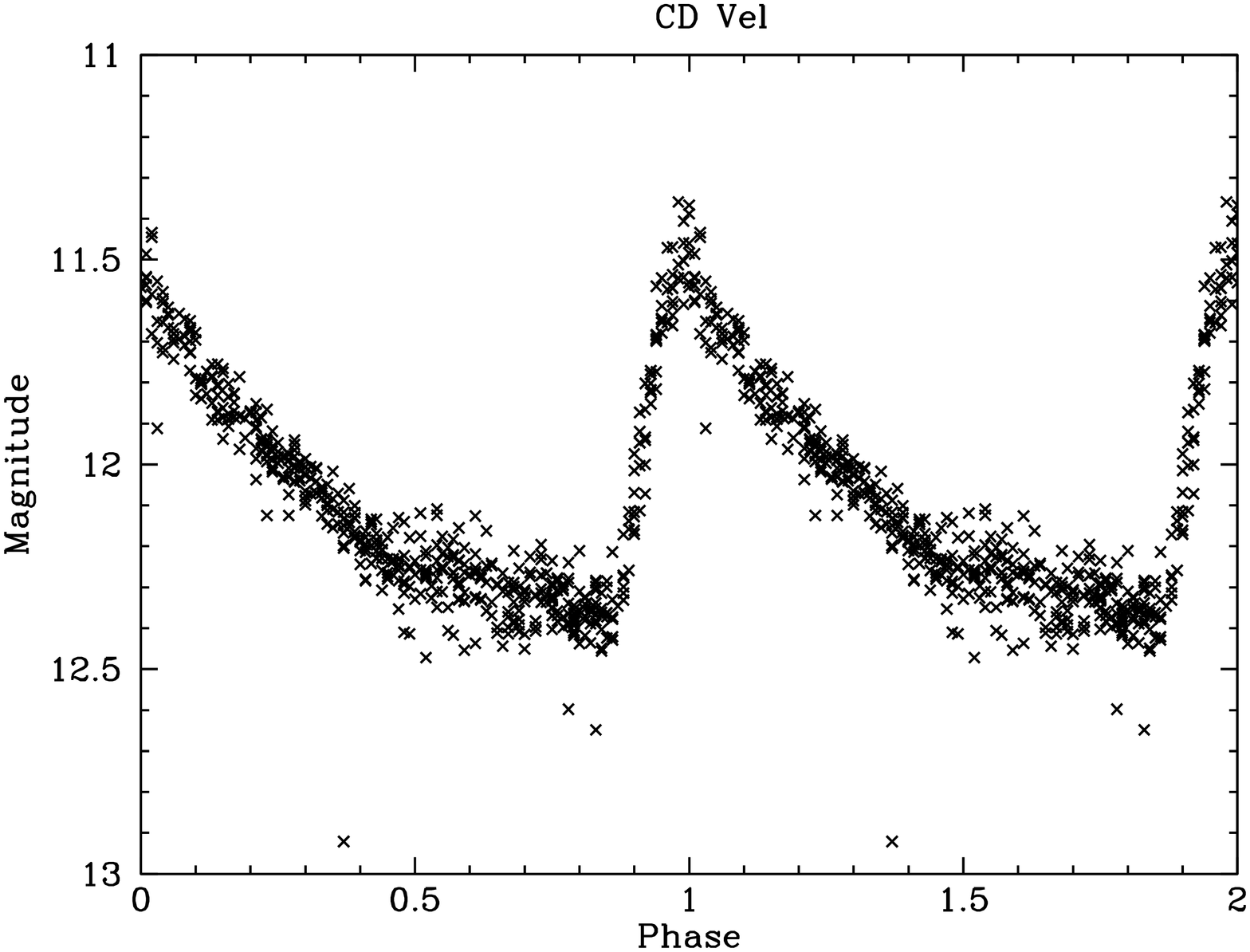}
\caption{Folded radial velocity curve and lightcurve of CD Vel by using our 
derived ephemeris for this star (Table~\ref{eph}). 
Top panel: Radial velocity vs pulsational phase for all of our spectra.  
The different symbols and colors represent different times of observations in HJD. 
The total numbers of observed spectra per cycle are listed in the parentheses.    
Bottom panel: The ASAS photometric lightcurve vs pulsational phase. 
The scatter of data points at a given phase for CD Vel and for the other program stars 
is highly related to the mean apparent brightness of the observed star. 
\label{RV-CDVel}}
\end{center}
\end{figure}

\begin{figure}
\begin{center}
\includegraphics[scale=0.5,angle=-90]{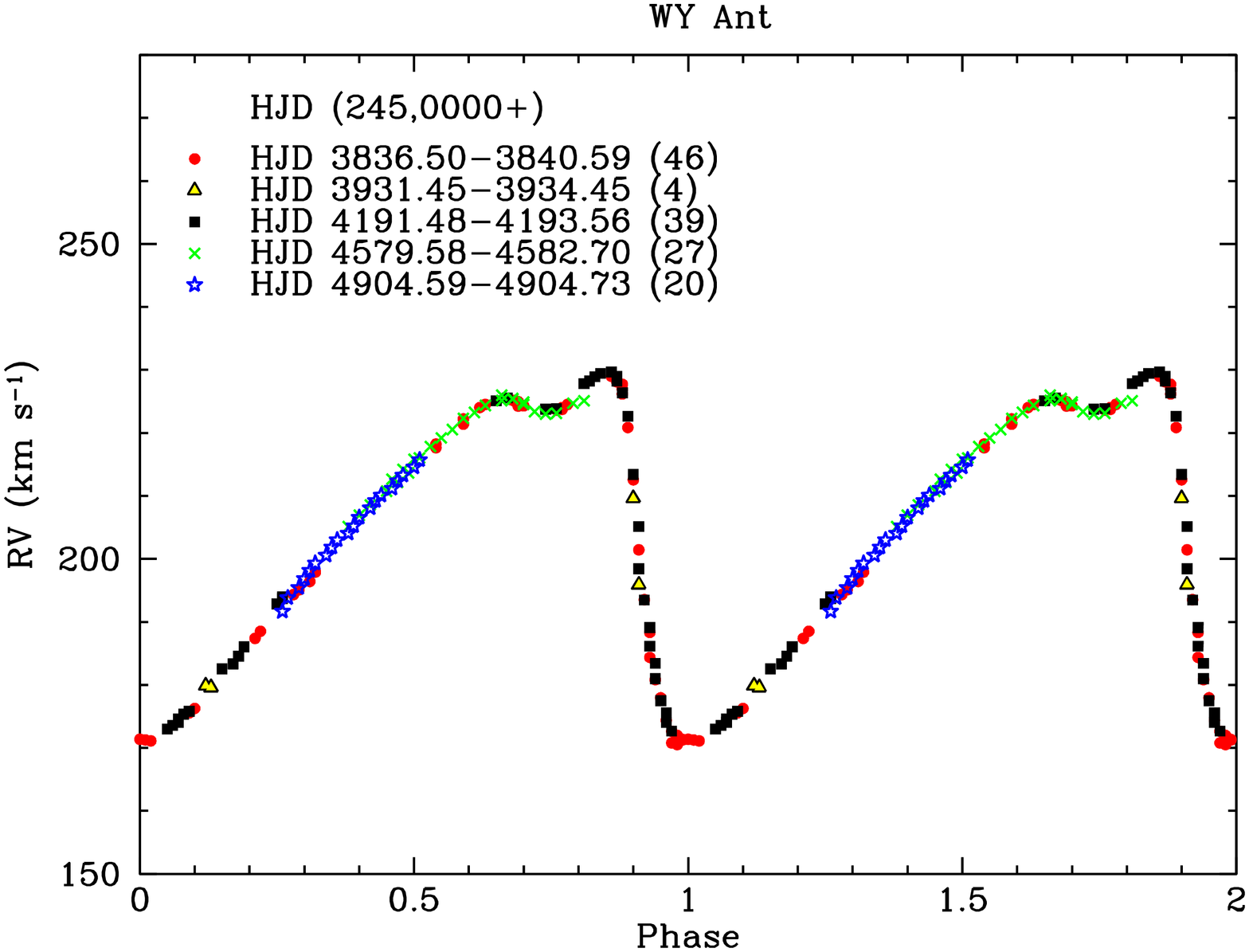}
\includegraphics[scale=0.5,angle=-90]{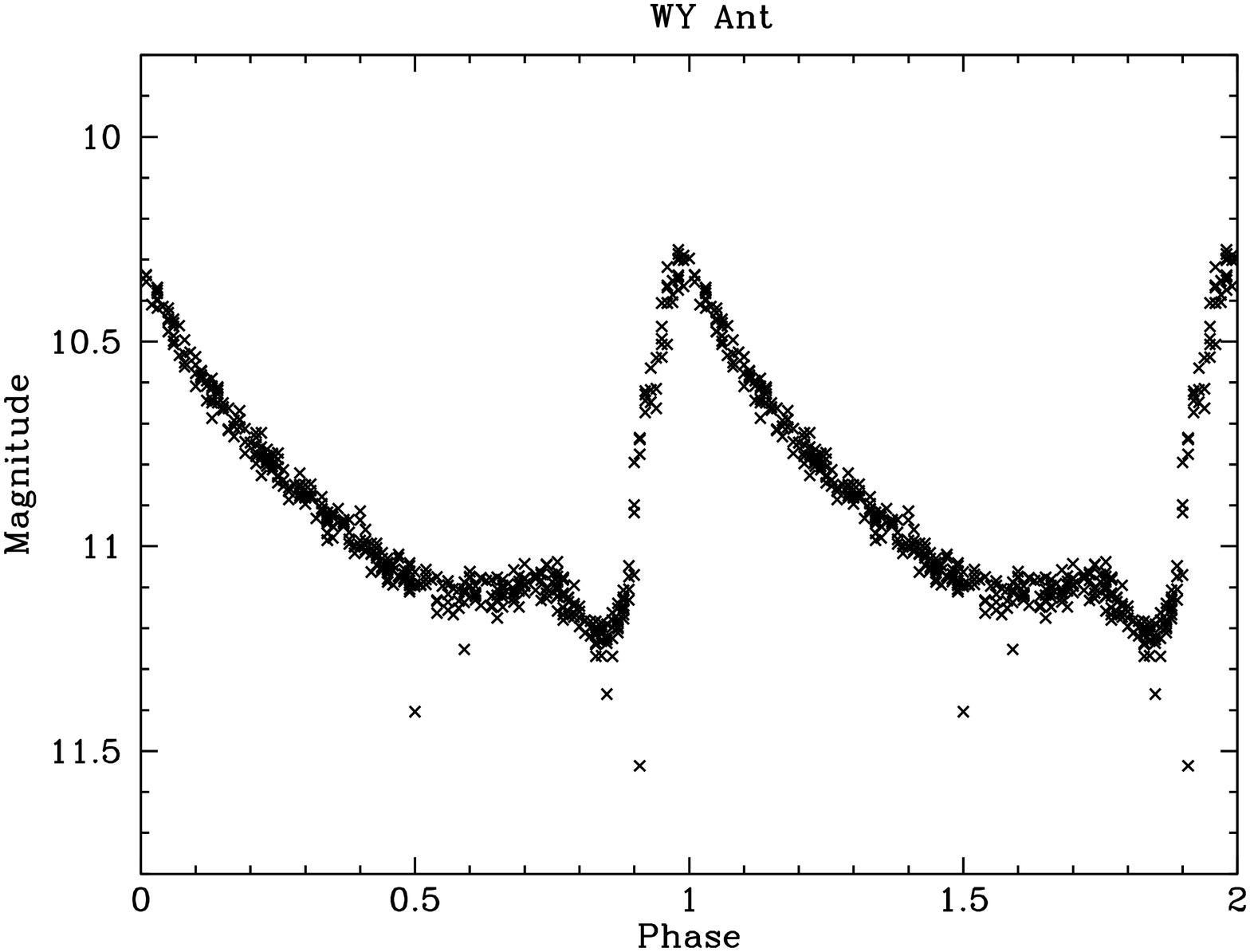}
\caption{Same as Figure~\ref{RV-CDVel}. \label{RV-WYAnt}}
\end{center}
\end{figure}

\begin{figure}
\begin{center}
\includegraphics[scale=0.5,angle=-90]{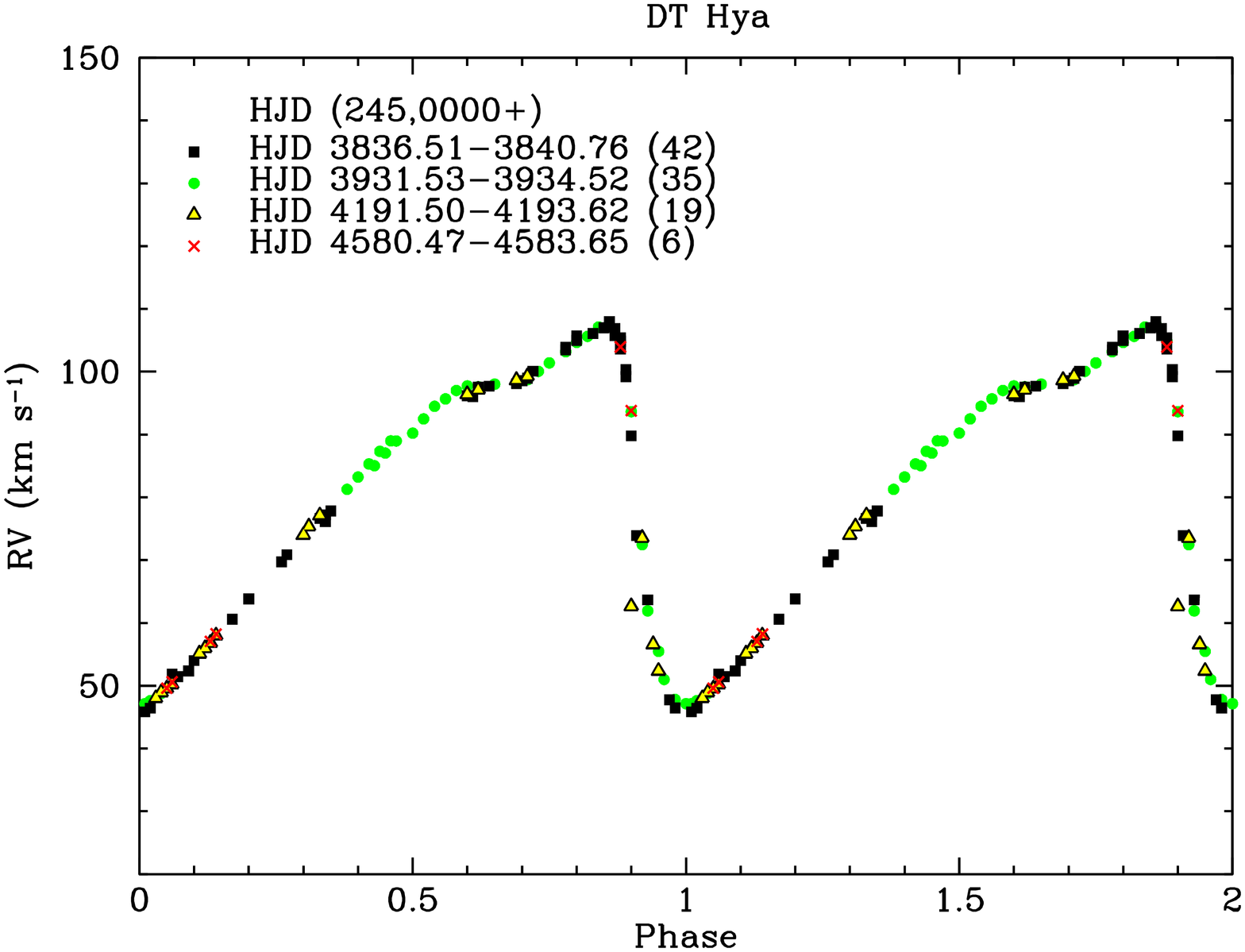}
\includegraphics[scale=0.5,angle=-90]{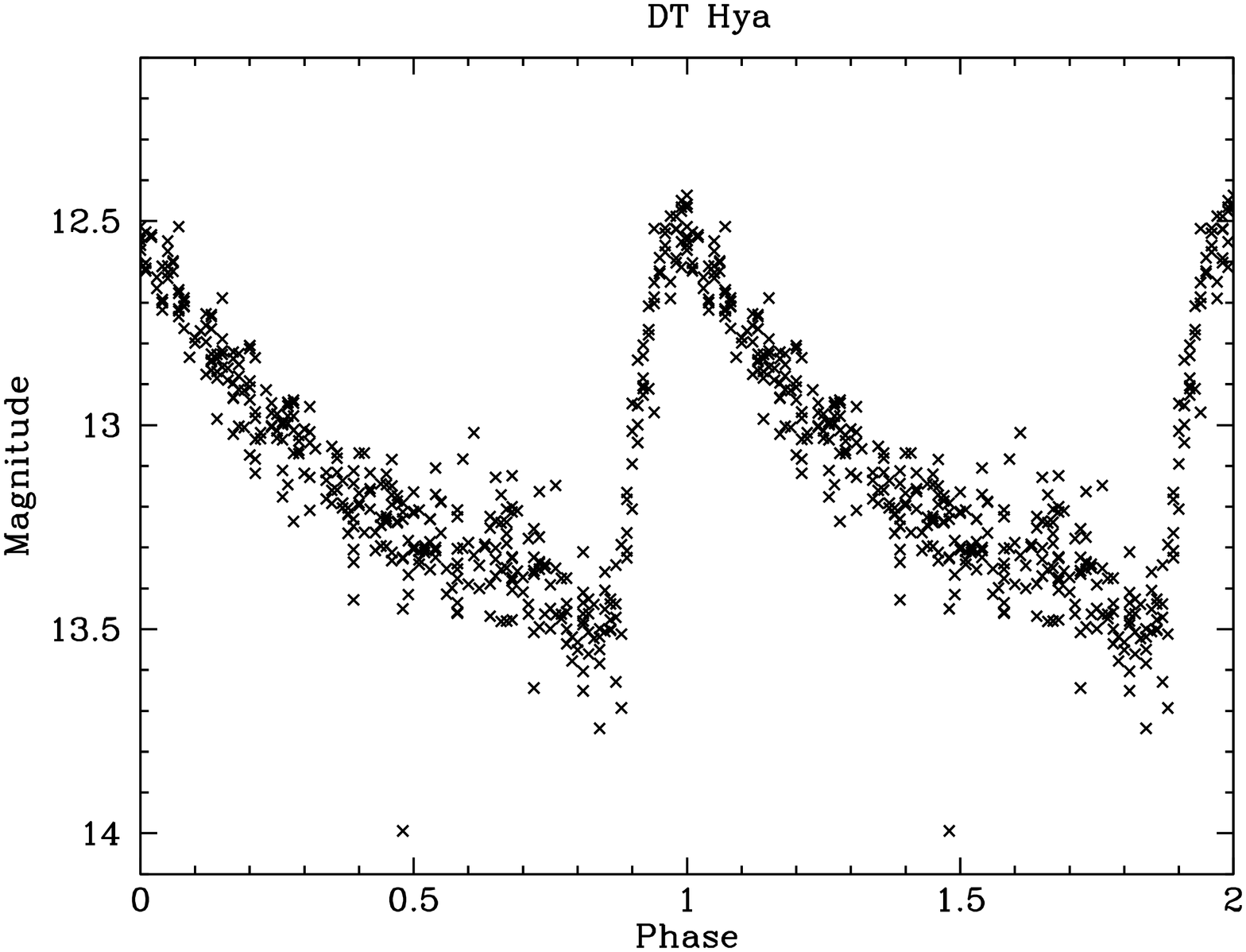}
\caption{Same as Figure~\ref{RV-CDVel}. \label{RV-DTHya}}
\end{center}
\end{figure}

\begin{figure}
\begin{center}
\includegraphics[scale=0.5,angle=-90]{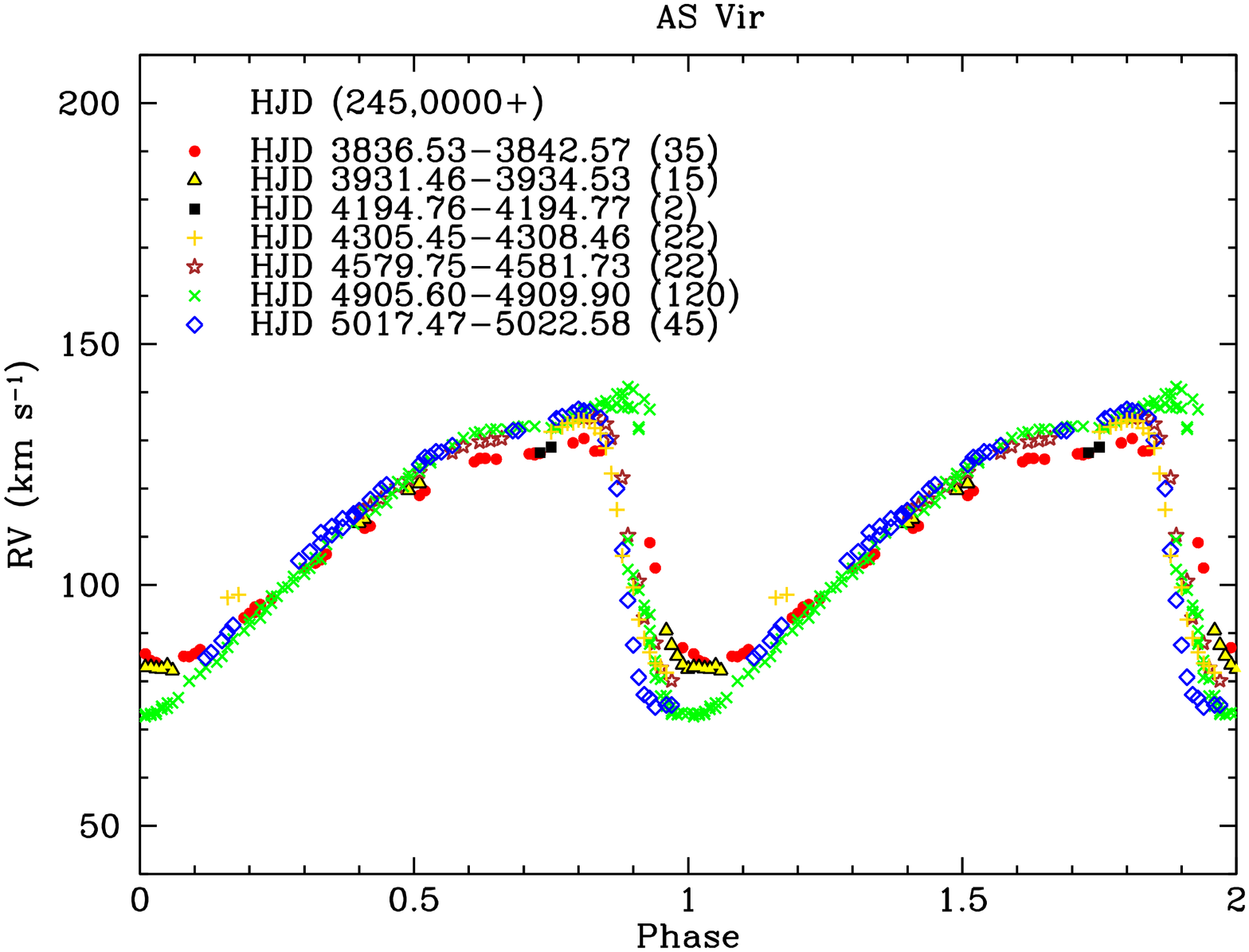}
\includegraphics[scale=0.5,angle=-90]{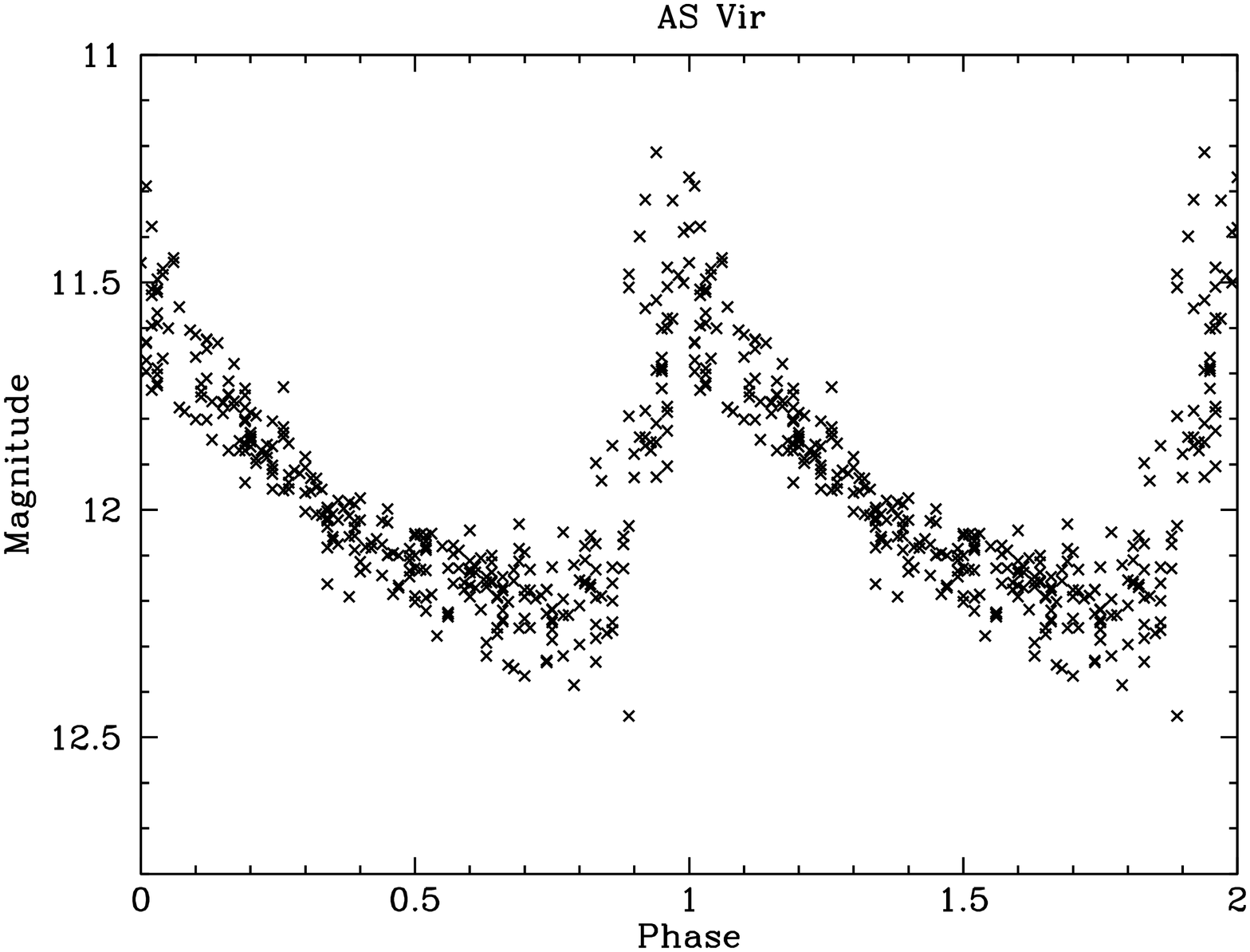}
\caption{Same as Figure~\ref{RV-CDVel}.\label{RV-ASVir}}
\end{center}
\end{figure}

\begin{figure}
\begin{center}
\includegraphics[scale=0.5,angle=-90]{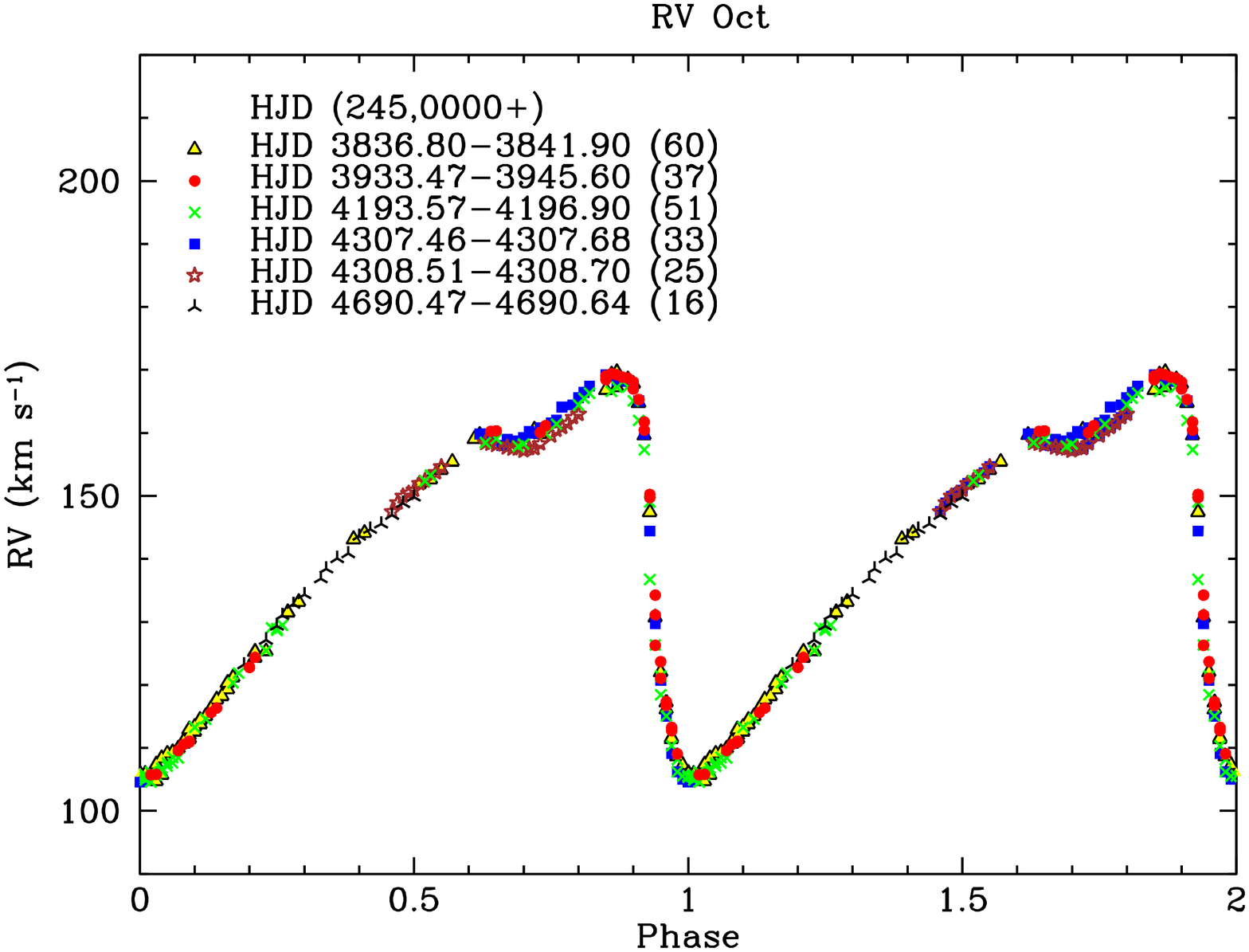}
\includegraphics[scale=0.5,angle=-90]{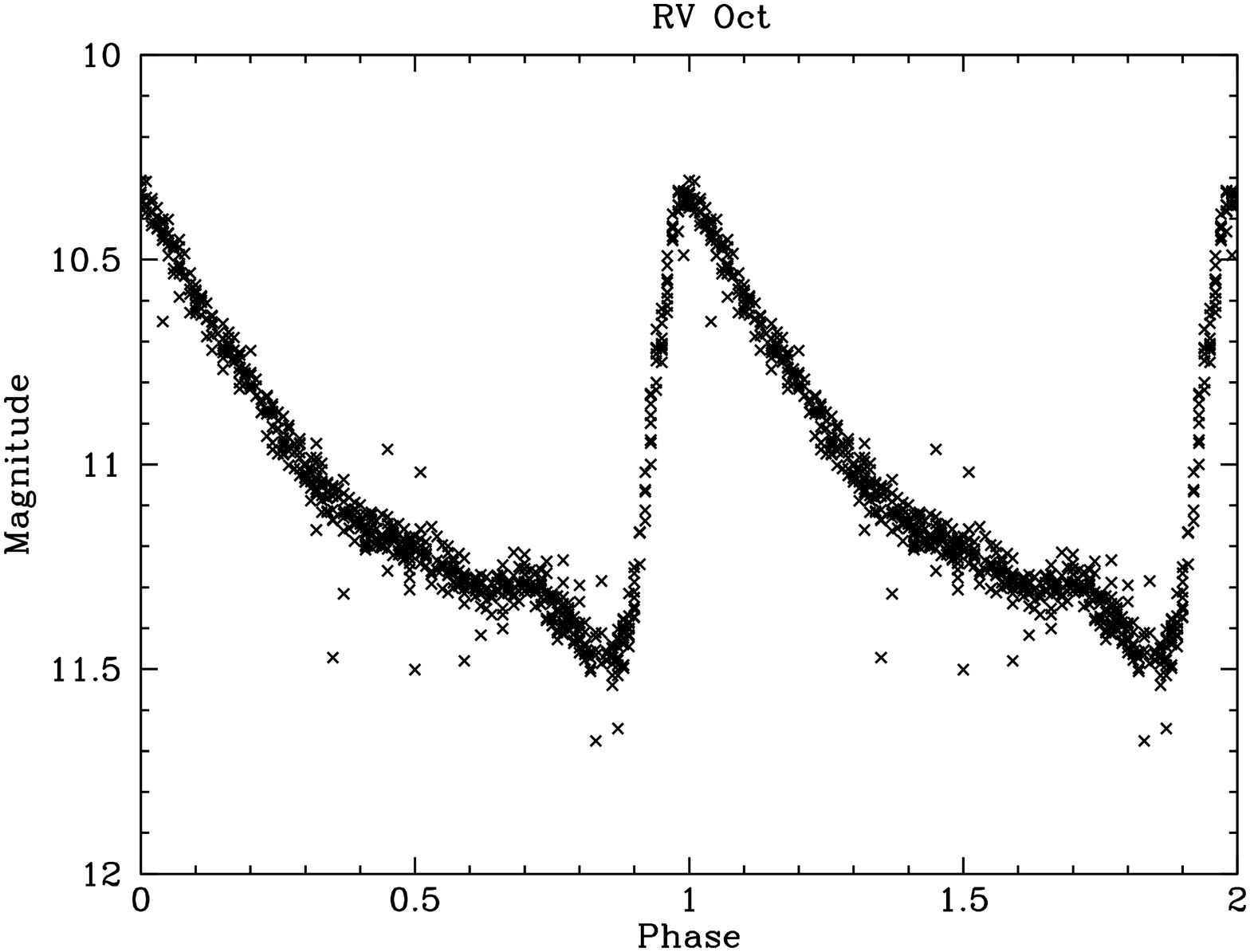}
\caption{Same as Figure~\ref{RV-CDVel}.\label{RV-RVOct}}
\end{center}
\end{figure}

\begin{figure}
\begin{center}
\includegraphics[scale=0.5,angle=-90]{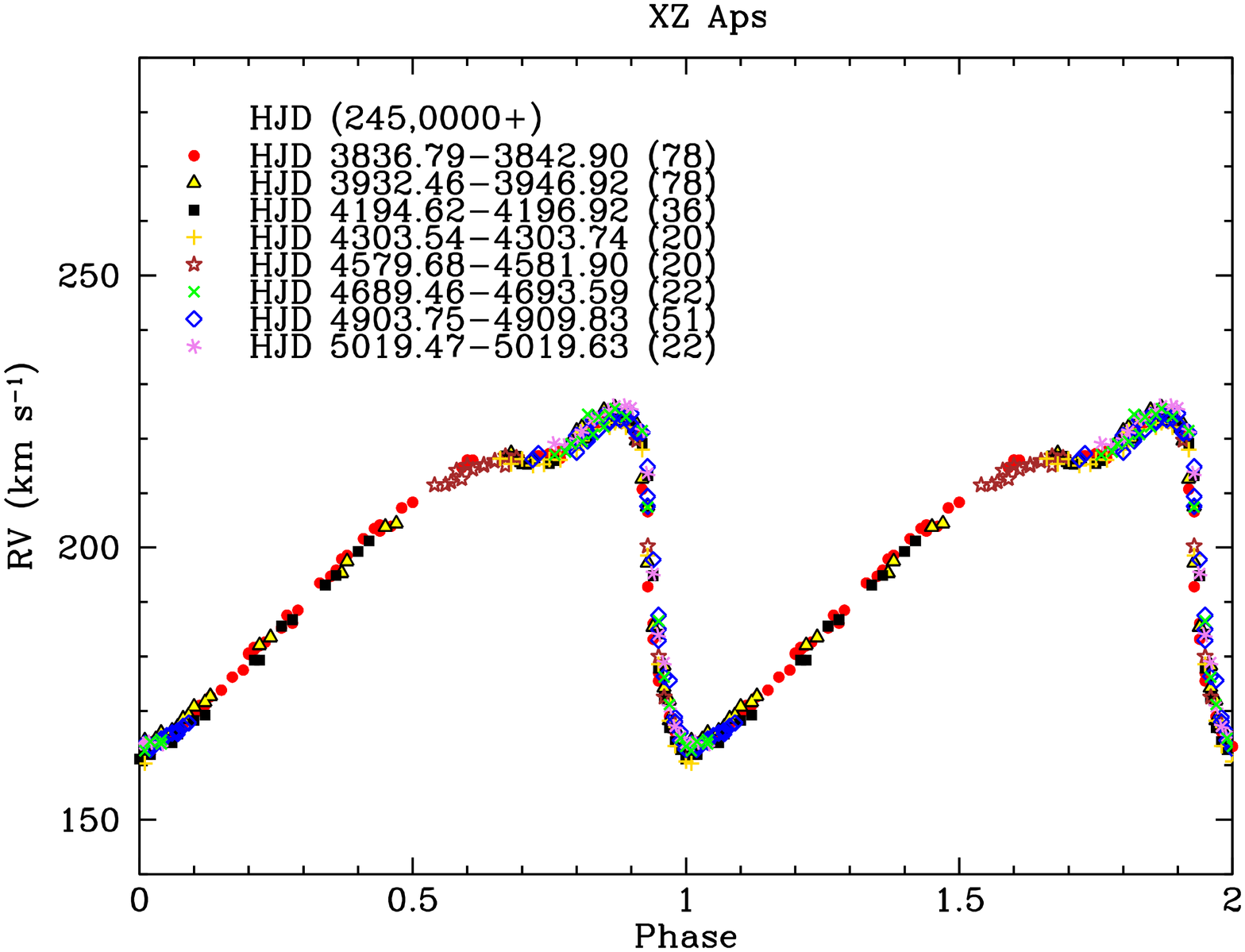}
\includegraphics[scale=0.5,angle=-90]{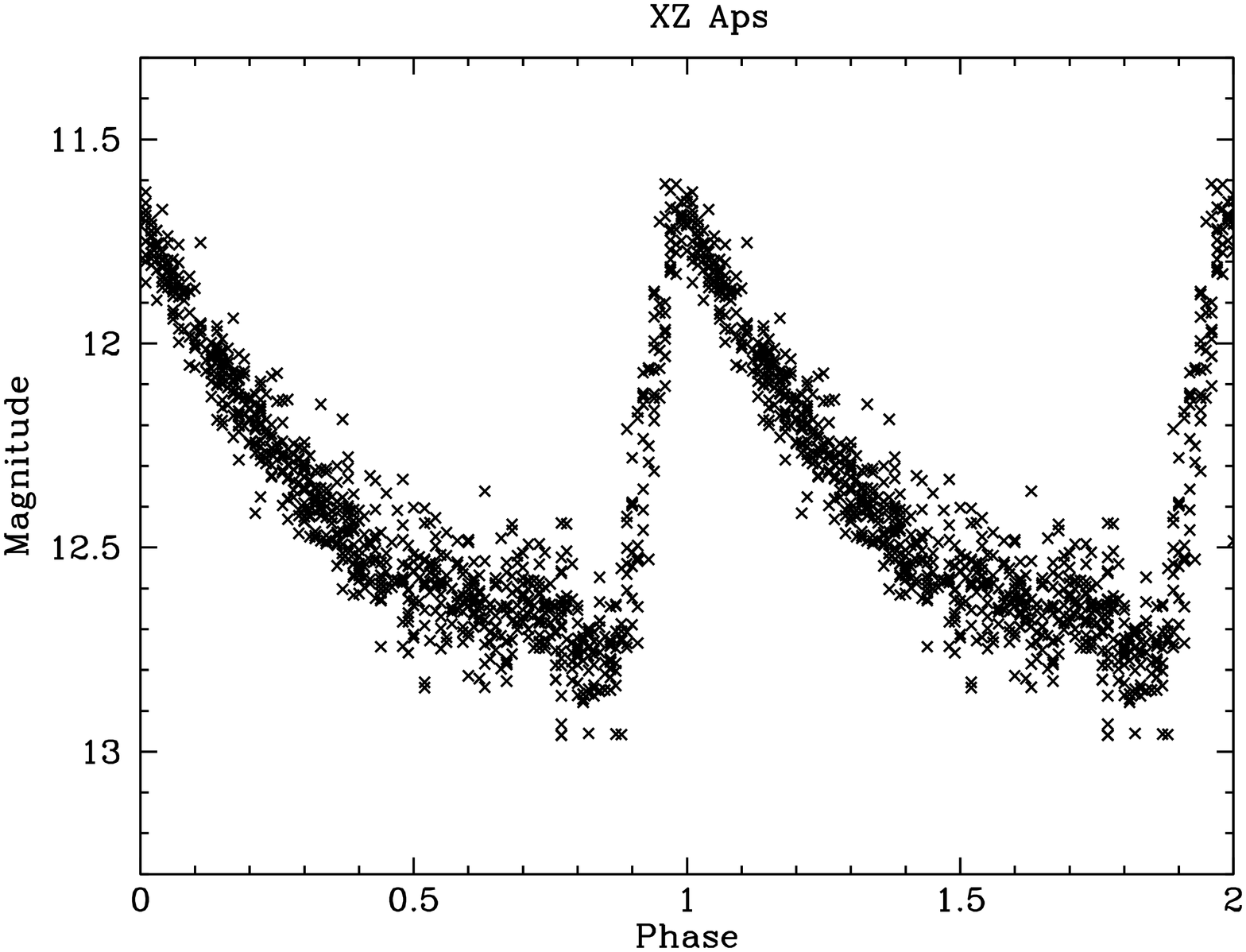}
\caption{Same as Figure~\ref{RV-CDVel}. \label{RV-XZAps}}
\end{center}
\end{figure}

\begin{figure}
\begin{center}
\includegraphics[scale=0.5,angle=-90]{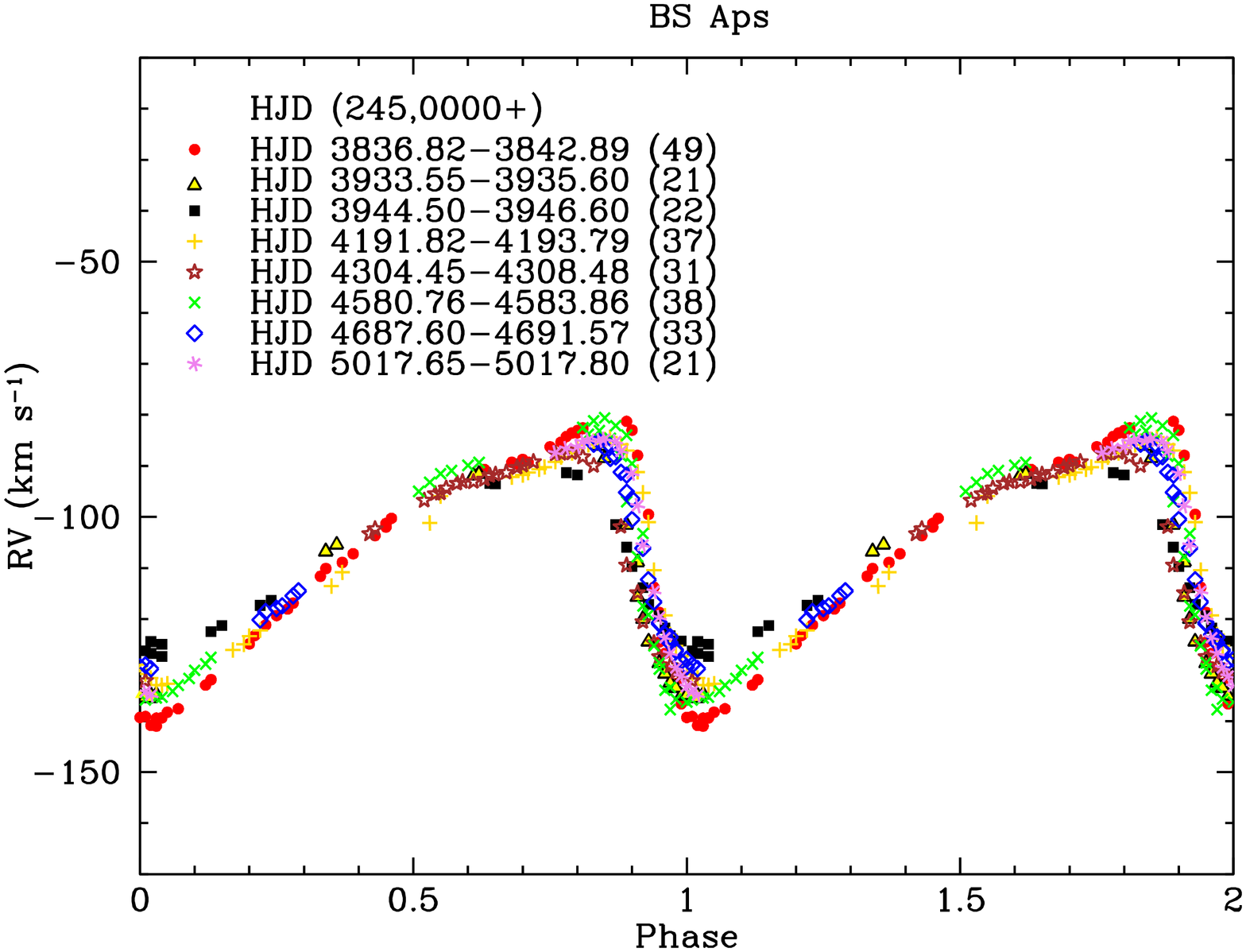}
\includegraphics[scale=0.5,angle=-90]{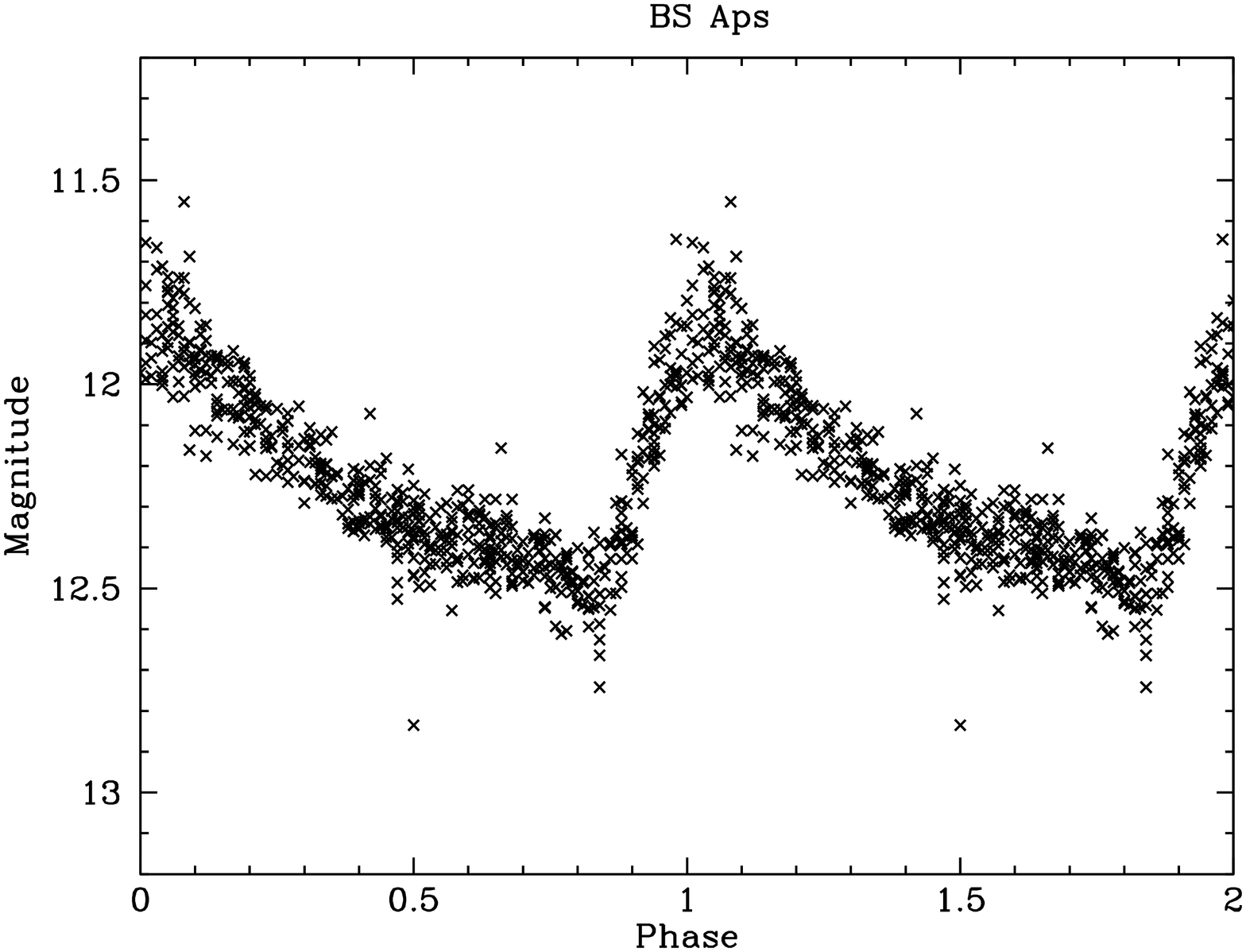}
\caption{Same as Figure~\ref{RV-CDVel}.
\label{RV-BSAps}}
\end{center}
\end{figure}

\begin{figure}
\begin{center}
\includegraphics[scale=0.5,angle=-90]{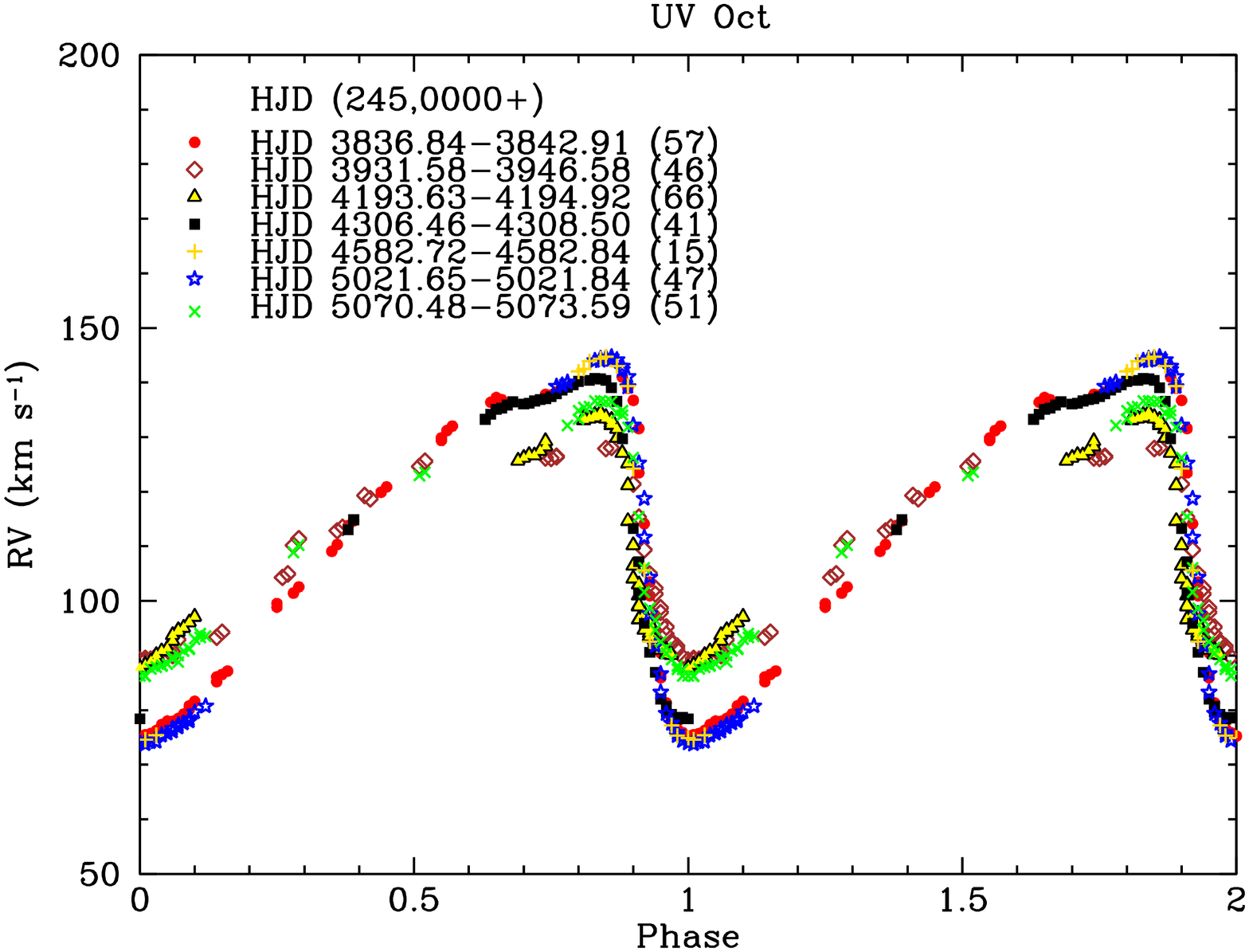}
\includegraphics[scale=0.5,angle=-90]{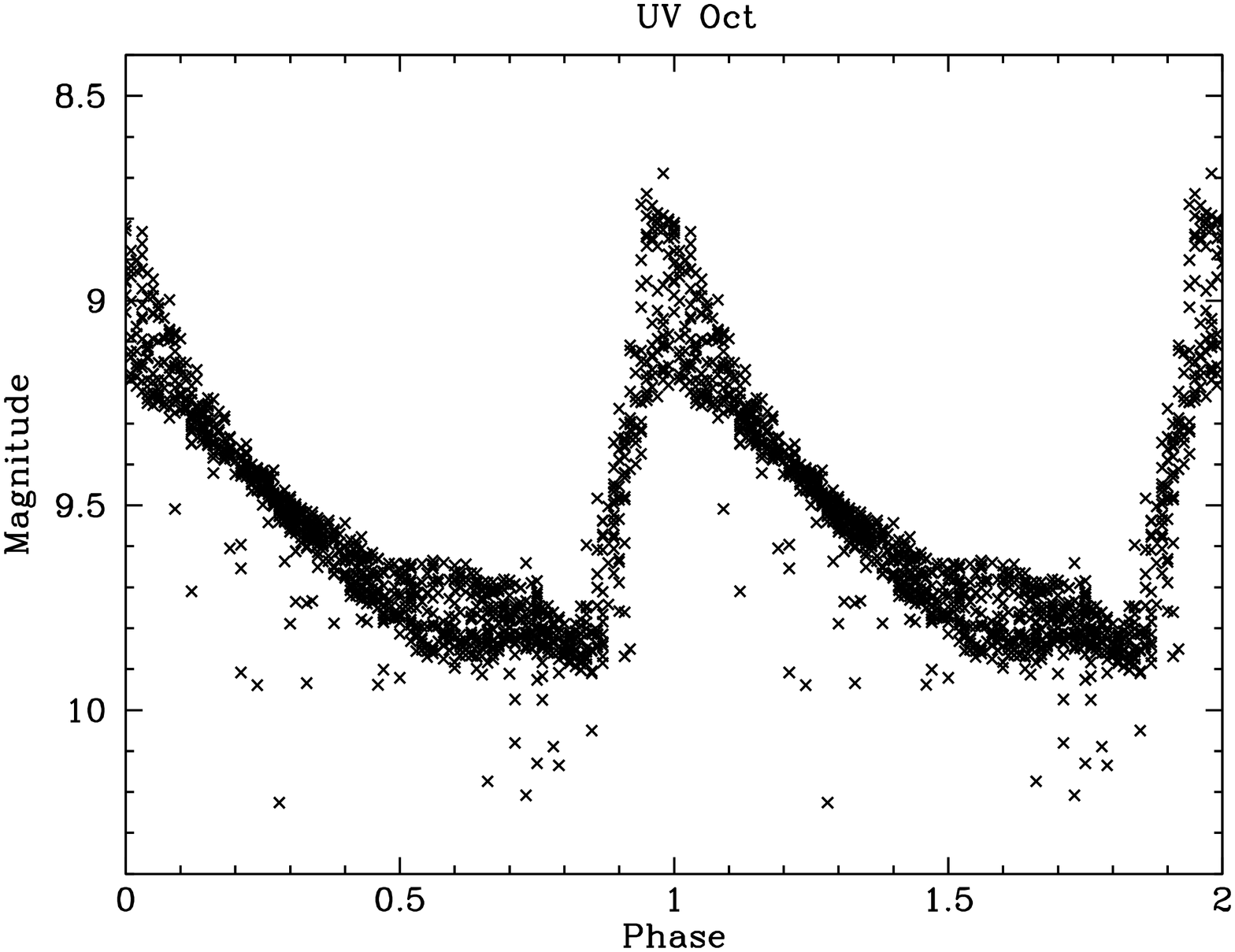}
\caption{Same as Figure~\ref{RV-CDVel}. \label{RV-UVOct}}
\end{center}
\end{figure}

\begin{figure}
\begin{center}
\includegraphics[scale=0.5,angle=-90]{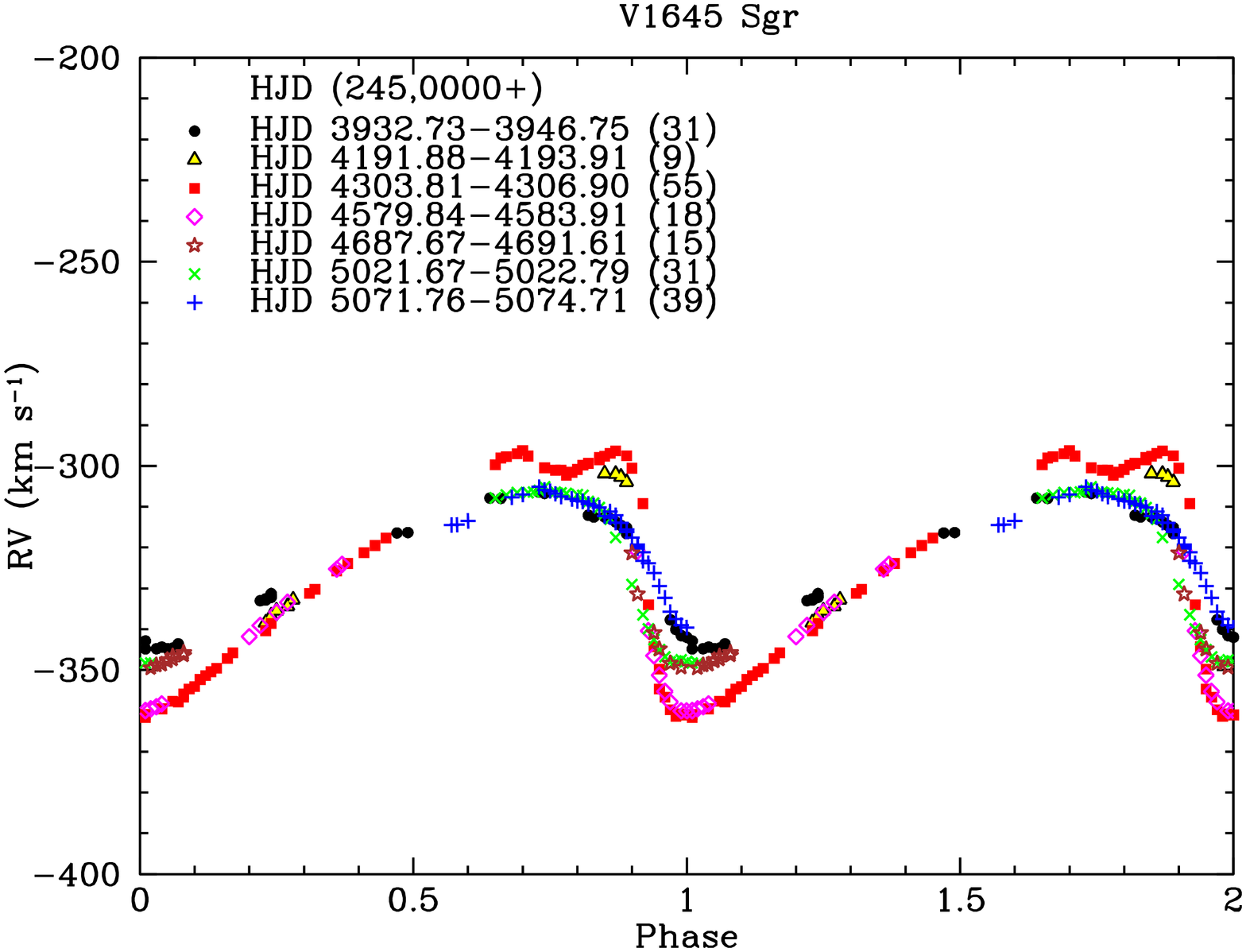}
\includegraphics[scale=0.5,angle=-90]{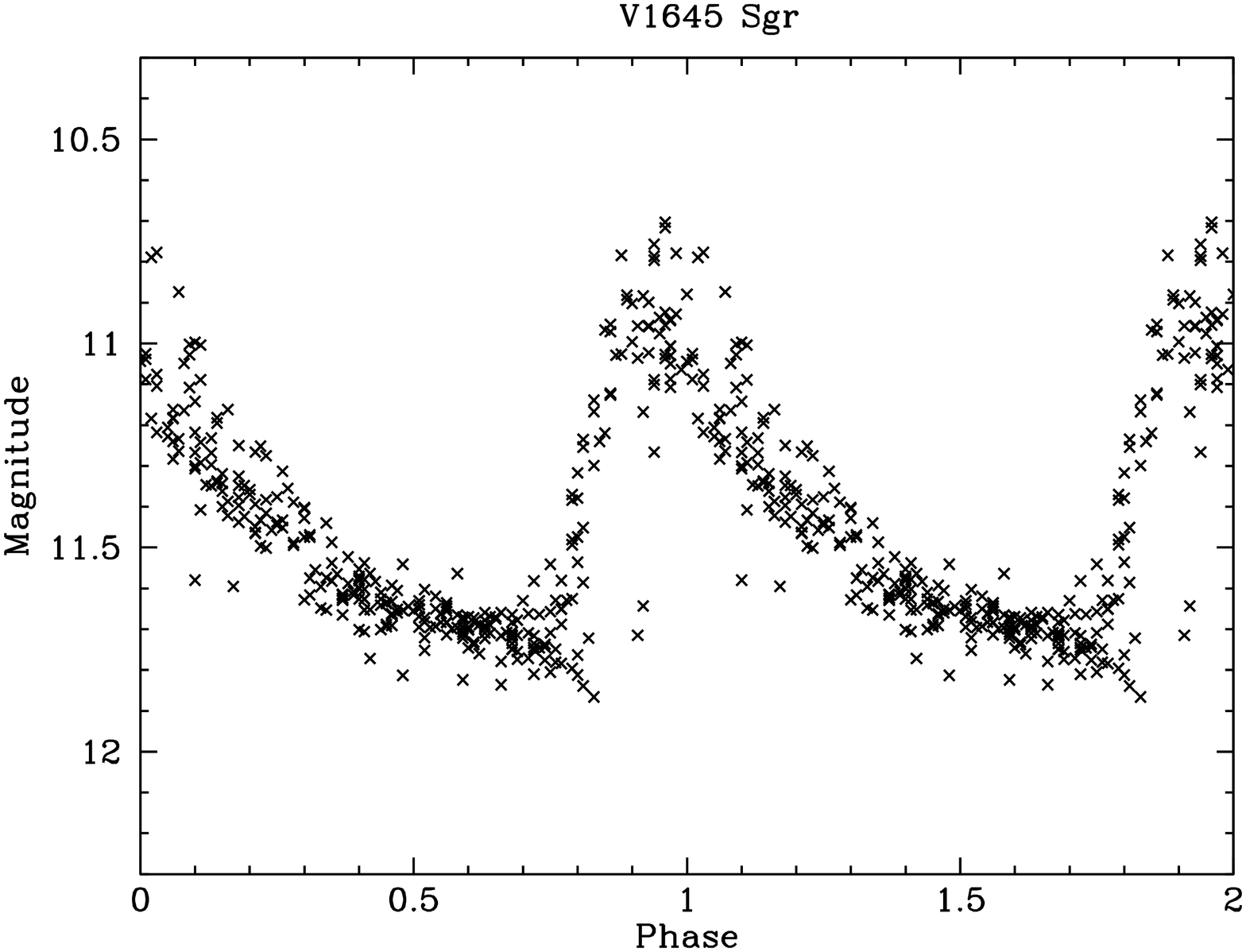}
\caption{Same as Figure~\ref{RV-CDVel}. \label{RV-VSgr}}
\end{center}
\end{figure}

\begin{figure}
\begin{center}
\includegraphics[scale=0.5,angle=-90]{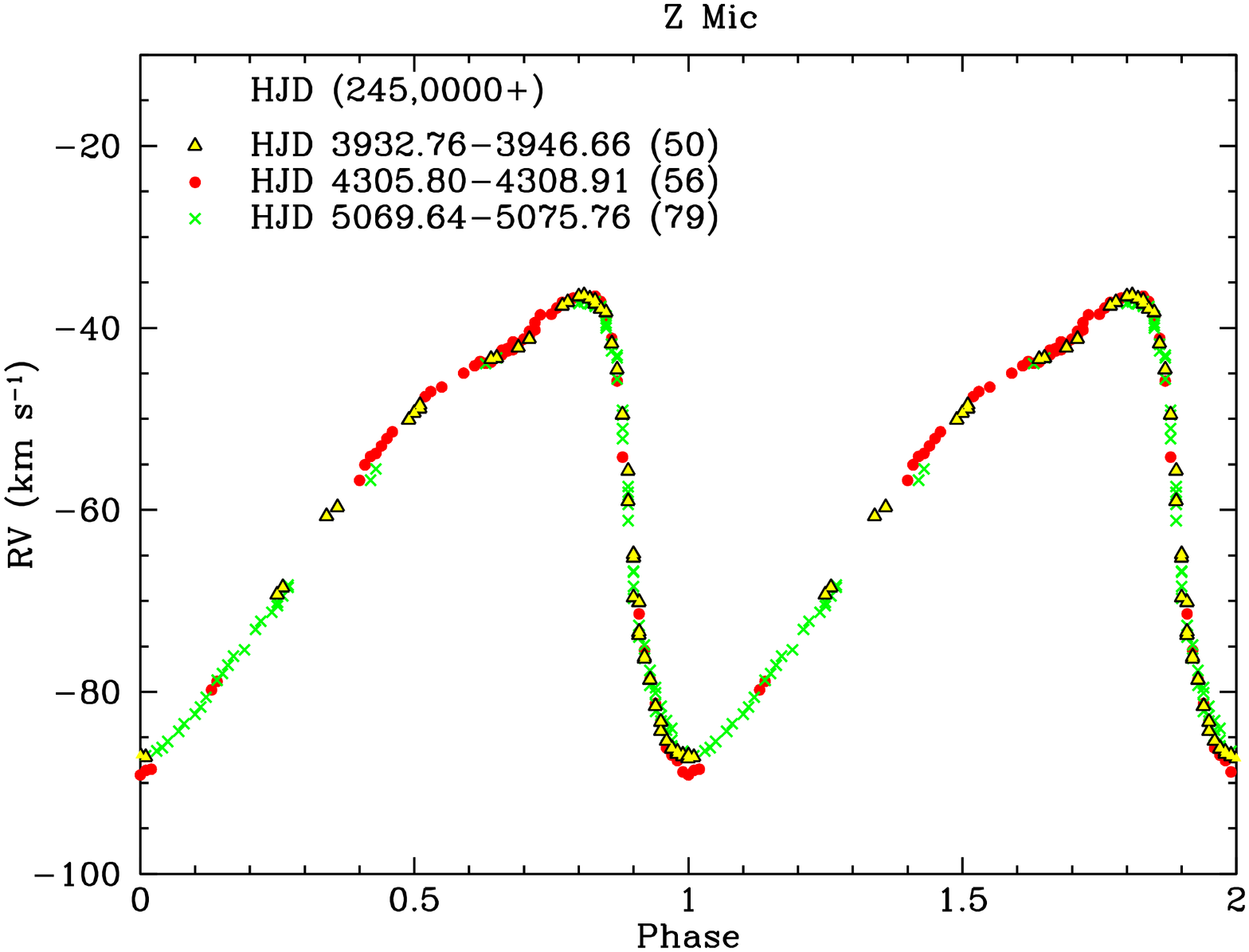}
\includegraphics[scale=0.5,angle=-90]{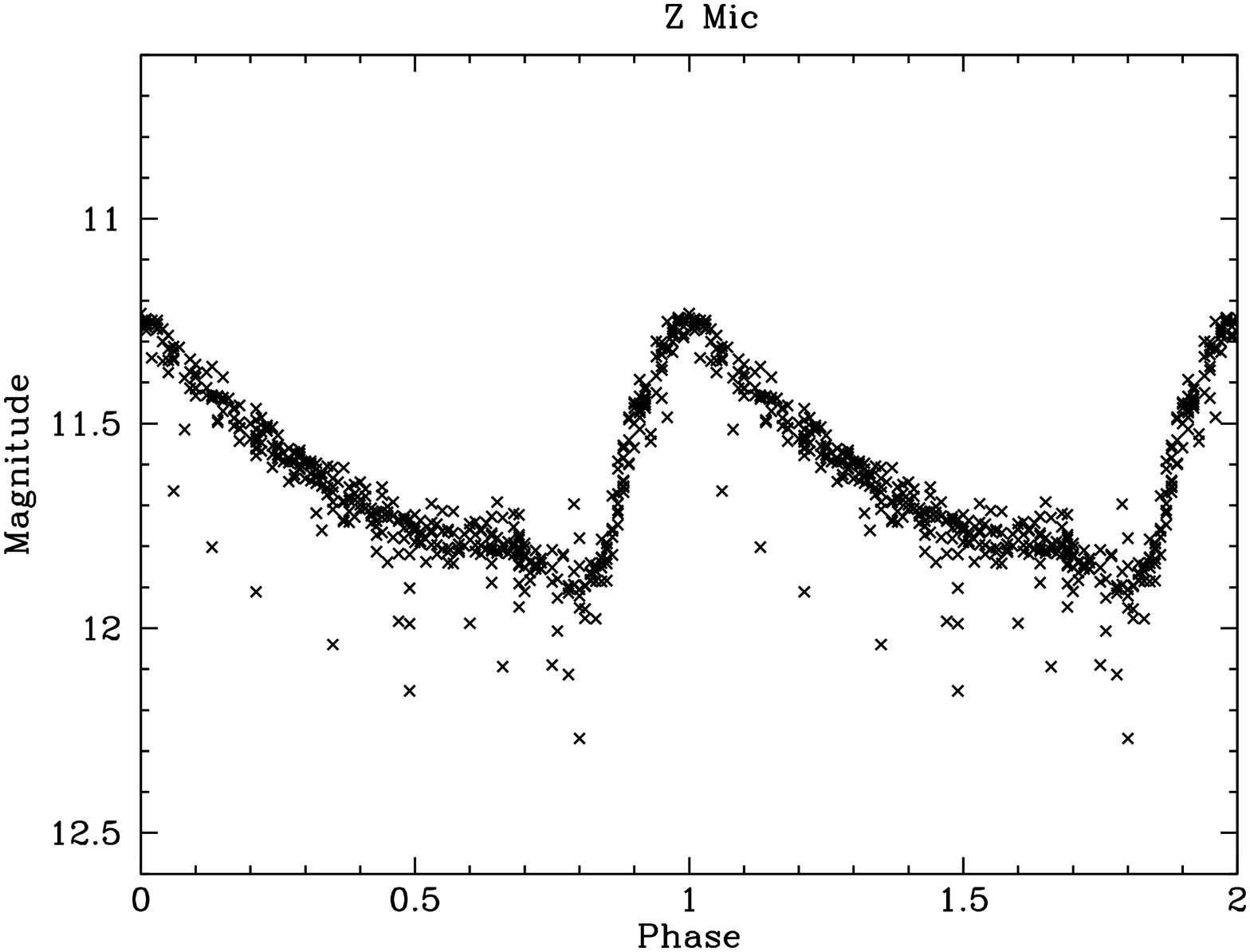}
\caption{Same as Figure~\ref{RV-CDVel}. \label{RV-ZMic}}
\end{center}
\end{figure}

\begin{figure}
\begin{center}
\includegraphics[scale=0.5,angle=-90]{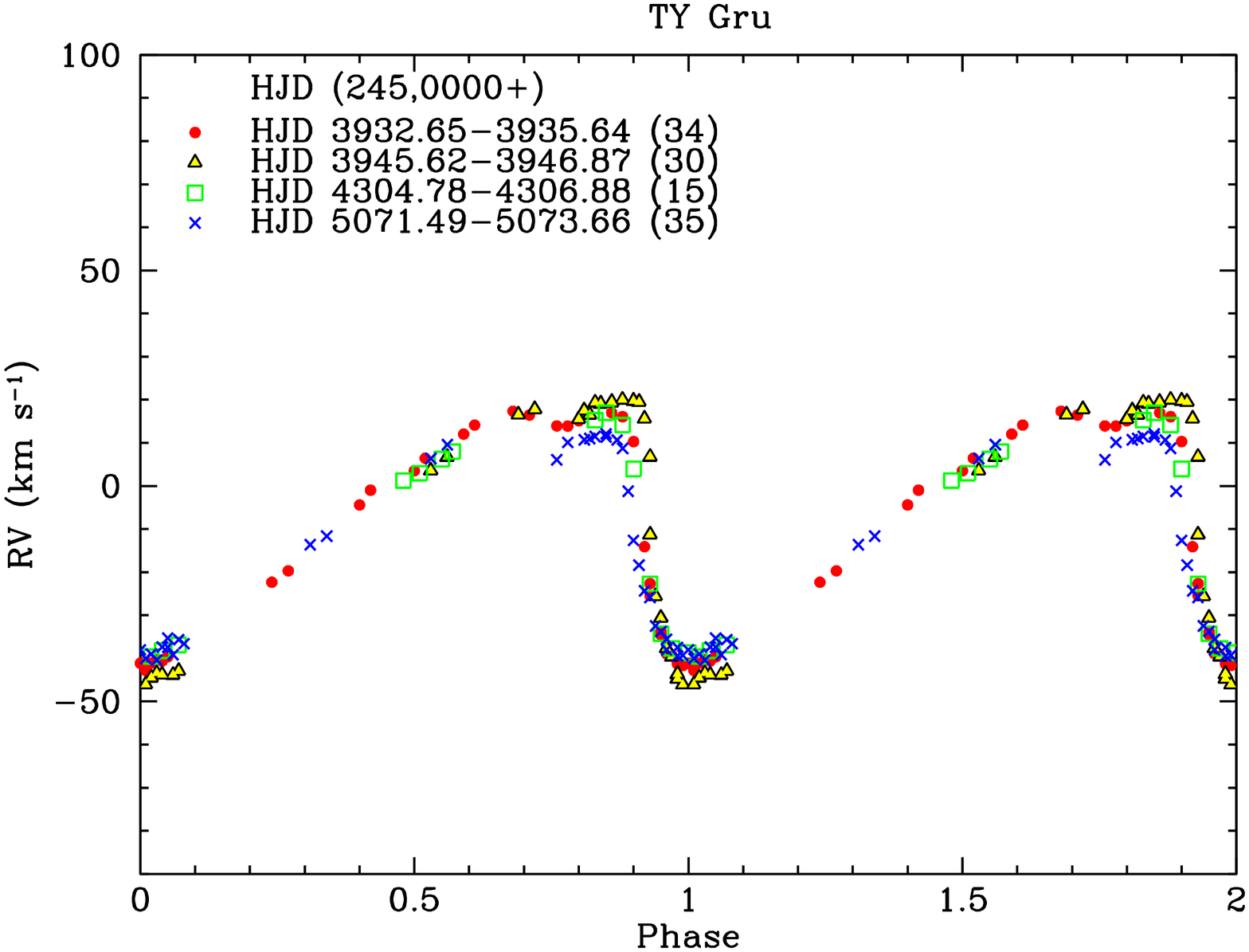}
\includegraphics[scale=0.5,angle=-90]{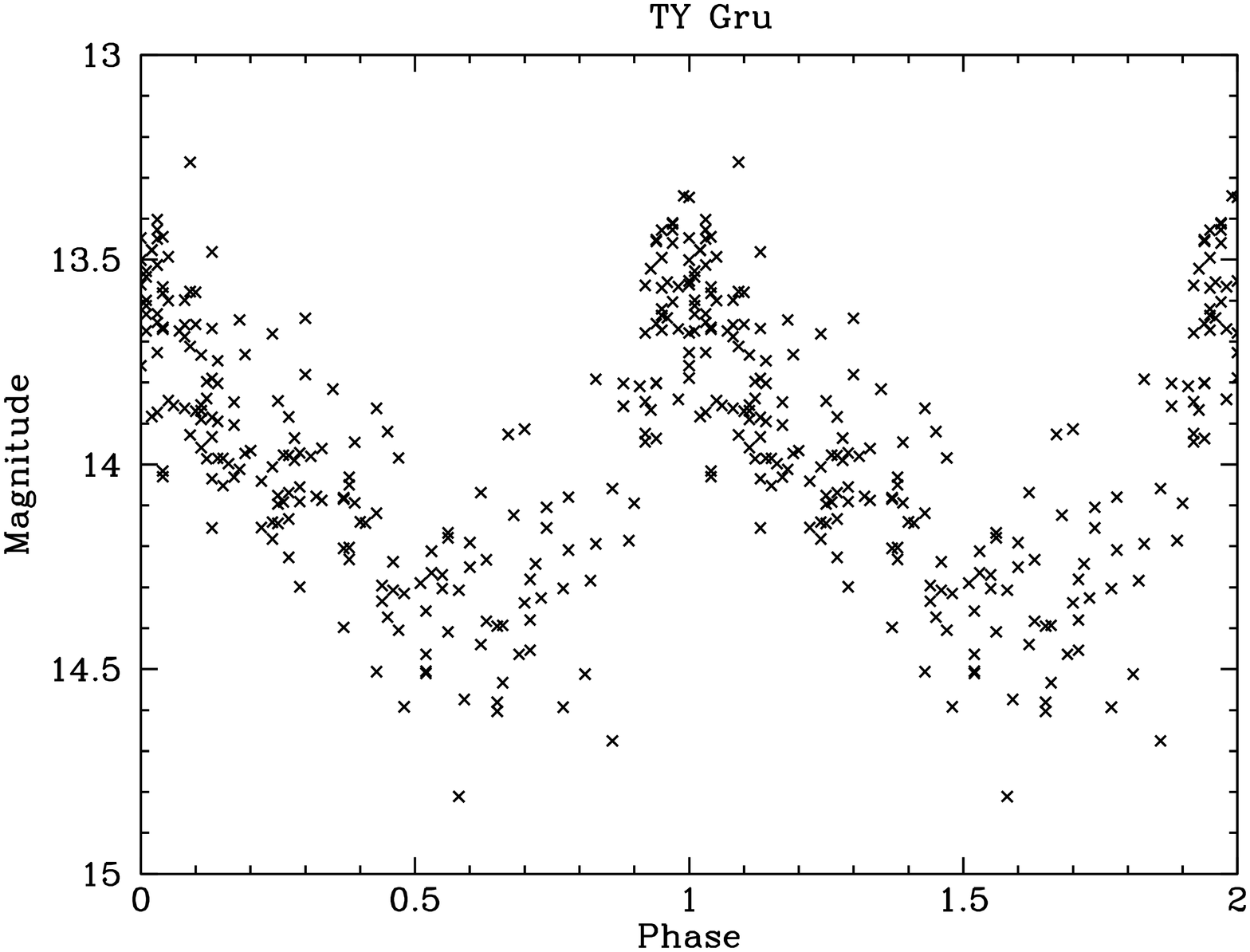}
\caption{Same as Figure~\ref{RV-CDVel}.\label{RV-TYGru}}
\end{center}
\end{figure}

In Figure~\ref{kvwdiagram}, we show the schematic 
diagram that determines the times of RV minima of our asymmetric RV curves. 
We refer the reader to \citet{KVW56} for the mathematical description of the method 
(for symmetric lightcurve only). 
In Table~\ref{eph}, we summarize the ephemerides of 11 field RRab stars. 
We tabulate epochs for the particular RV minima used to derive them.  
The table also gives the range of data in HJD that are associated with the 
corresponding $T_{0}$ and pulsational periods.  
We present the folded RV curves and ASAS lightcurves with our derived 
ephemerides in 
top and bottom panels of Figure~\ref{RV-CDVel}--\ref{RV-TYGru}. 
The figures are arranged by ascending right ascension.

\acknowledgments
B.-Q. For acknowledges travel assistance from a SigmaXi grant-in-aid. 
This research was supported by U.S. National Science Foundation grants
AST-0908978.


\bibliography{ref}

\begin{thebibliography}{18}
\expandafter\ifx\csname natexlab\endcsname\relax\def\natexlab#1{#1}\fi

\bibitem[{{Beers} {et~al.}(1992){Beers}, {Preston}, \& {Shectman}}]{BPS92}
{Beers}, T.~C., {Preston}, G.~W., \& {Shectman}, S.~A. 1992, \aj, 103, 1987

\bibitem[{{For} \& {Sneden}(2010)}]{FS10}
{For}, B., \& {Sneden}, C. 2010, \aj, 140, 1694

\bibitem[{{Ivezi{\'c}} {et~al.}(2004){Ivezi{\'c}}, {Lupton}, {Schlegel},
  {Smol{\v c}i{\'c}}, {Johnston}, {Gunn}, {Knapp}, {Strauss}, {Rockosi}, \&
  {The SDSS Collaboration}}]{Ivezic04}
{Ivezi{\'c}}, {\v Z}., {et~al.} 2004, in Astronomical Society of the Pacific
  Conference Series, Vol. 327, Satellites and Tidal Streams, ed. {F.~Prada,
  D.~Martinez Delgado, \& T.~J.~Mahoney}, 104--+

\bibitem[{{Kov{\'a}cs} {et~al.}(2002){Kov{\'a}cs}, {Zucker}, \&
  {Mazeh}}]{Kovacs02}
{Kov{\'a}cs}, G., {Zucker}, S., \& {Mazeh}, T. 2002, \aap, 391, 369

\bibitem[{{Kwee} \& {van Woerden}(1956)}]{KVW56}
{Kwee}, K.~K., \& {van Woerden}, H. 1956, \bain, 12, 327

\bibitem[{{Majewski} {et~al.}(2003){Majewski}, {Skrutskie}, {Weinberg}, \&
  {Ostheimer}}]{Majewski03}
{Majewski}, S.~R., {Skrutskie}, M.~F., {Weinberg}, M.~D., \& {Ostheimer}, J.~C.
  2003, \apj, 599, 1082

\bibitem[{{Pojmanski}(2002)}]{Pojmanski02}
{Pojmanski}, G. 2002, \actaa, 52, 397

\bibitem[{{Preston}(1959)}]{Preston59}
{Preston}, G.~W. 1959, \apj, 130, 507

\bibitem[{{Preston}(2009)}]{Preston09}
---. 2009, \aap, 507, 1621

\bibitem[{{Preston}(2011)}]{Preston11}
---. 2011, \aj, 141, 6

\bibitem[{{Preston} {et~al.}(1991){Preston}, {Shectman}, \&
  {Beers}}]{Preston91}
{Preston}, G.~W., {Shectman}, S.~A., \& {Beers}, T.~C. 1991, \apjs, 76, 1001

\bibitem[{{Preston} \& {Sneden}(2000)}]{PS00}
{Preston}, G.~W., \& {Sneden}, C. 2000, \aj, 120, 1014

\bibitem[{{Preston} {et~al.}(2006{\natexlab{a}}){Preston}, {Sneden},
  {Thompson}, {Shectman}, \& {Burley}}]{PrestonRHB06}
{Preston}, G.~W., {Sneden}, C., {Thompson}, I.~B., {Shectman}, S.~A., \&
  {Burley}, G.~S. 2006{\natexlab{a}}, \aj, 132, 85

\bibitem[{{Preston} {et~al.}(2006{\natexlab{b}}){Preston}, {Thompson},
  {Sneden}, {Stachowski}, \& {Shectman}}]{Preston06}
{Preston}, G.~W., {Thompson}, I.~B., {Sneden}, C., {Stachowski}, G., \&
  {Shectman}, S.~A. 2006{\natexlab{b}}, \aj, 132, 1714

\bibitem[{{Scargle}(1982)}]{Scargle82}
{Scargle}, J.~D. 1982, \apj, 263, 835

\bibitem[{{Simon}(1966)}]{Simon66}
{Simon}, G.~W. 1966, \aj, 71, 190

\bibitem[{{Szczygie{\l}} \& {Fabrycky}(2007)}]{SF07}
{Szczygie{\l}}, D.~M., \& {Fabrycky}, D.~C. 2007, \mnras, 377, 1263

\bibitem[{{Vivas} {et~al.}(2004){Vivas}, {Zinn}, {Abad}, {Andrews}, {Bailyn},
  {Baltay}, {Bongiovanni}, {Brice{\~n}o}, {Bruzual}, {Coppi}, {Della Prugna},
  {Ellman}, {Ferr{\'{\i}}n}, {Gebhard}, {Girard}, {Hernandez}, {Herrera},
  {Honeycutt}, {Magris}, {Mufson}, {Musser}, {Naranjo}, {Rabinowitz},
  {Rengstorf}, {Rosenzweig}, {S{\'a}nchez}, {S{\'a}nchez}, {Schaefer},
  {Schenner}, {Snyder}, {Sofia}, {Stock}, {van Altena}, {Vicente}, \&
  {Vieira}}]{Vivas04}
{Vivas}, A.~K., {et~al.} 2004, \aj, 127, 1158

\end{thebibliography}

\clearpage

\clearpage

\clearpage

\begin{deluxetable}{lccrrc}
\tabletypesize{\scriptsize}
\tablewidth{0pt}
\tablecolumns{6}
\tablecaption{Program stars.\label{targets}}
\tablehead{
\colhead{Star} & 
\colhead{R.A.(J2000)}  & 
\colhead{Decl.(J2000)}  & 
\colhead{$V_{\rm max}$\tablenotemark{a}}  &
\colhead{$V_{\rm amp}$\tablenotemark{a}}  & 
\colhead{Note}  \\ 
\colhead{} &
\colhead{(hr m s)} &
\colhead{($\degr\,\arcmin\,\arcsec$)}   & 
\colhead{(mag)}  &
\colhead{(mag)}  & 
\colhead{}
}
\startdata
CD Vel	&	09 44 38.24	&	$-$45 52 37.2	& 11.66	&	0.87        &	Blazhko\tablenotemark{c}	\\
WY Ant	&	10 16 04.95	&	$-$29 43 42.4	& 10.37	&	0.85        &	\nodata	\\
DT Hya	&	11 54 00.18	&	$-$31 15 40.0	& 12.53	&	0.98        &	\nodata	\\
AS Vir	&	12 52 45.86	&	$-$10 15 36.4	& 11.66	&	0.72        &	Blazhko	\\
RV Oct	&	13 46 31.75	&	$-$84 24 06.4	& 10.53	&	1.13        &	Blazhko	\\
XZ Aps	&	14 52 05.43	&	$-$79 40 46.6	& 11.94	&	1.1         &	\nodata	\\
BS Aps	&	16 20 51.51	&	$-$71 40 15.8	& 11.9	&	0.68        &	Blazhko	\\
UV Oct	&	16 32 25.53	&	$-$83 54 10.5	& 9.19	&	0.82        &	Blazhko	\\
V1645 Sgr&      20 20 44.47     &	$-$41 07 05.7	& 10.99	&	0.84        &	Blazhko	\\
Z Mic	&	21 16 22.71	&	$-$30 17 03.1	& 11.32	&	0.64        &	Blazhko \\
TY Gru	&	22 16 39.42	&	$-$39 56 18.0	& 13.6\tablenotemark{d}	&	0.9\tablenotemark{d}         &	Blazhko	
\enddata
\tablenotetext{a}{Maximum light in $V$ magnitude from ASAS.}
\tablenotetext{b}{Pulsational amplitude in $V$-band from ASAS. }
\tablenotetext{c}{\citet{SF07}.}
\tablenotetext{d}{Values extracted from \citet{Preston06}.}
\end{deluxetable}

\clearpage
\begin{deluxetable}{clccrrr}
\tabletypesize{\scriptsize}
\tablewidth{0pt}
\tablecolumns{7}
\tablecaption{Basic Information and Observing Log of Standard Stars\label{standard}}
\tablehead{
\colhead{Star} &
\colhead{Spectral Type} &
\colhead{R.A.} &
\colhead{Decl.} &
\colhead{$V$} &
\colhead{UT Date} &
\colhead{$N_{\rm exp}$} \\
\colhead{} &
\colhead{} &
\colhead{(h~m~s)} &
\colhead{($\degr~\arcmin~\arcsec$)} &
\colhead{(mag)} &
\colhead{} &
\colhead{}
}
\startdata
HD~142629	&	A3~V	&	15 56 53.498	&	$-$33 57 58.08	&	5.095	&	08,09 Aug 2008	&	2,2	\\
HD~135153	&	F1~III	&	15 14 37.319	&	$-$31 31 08.84	&	4.924	&	09 Aug 2008	&	3	\\
HD~144880	&	F7~V	&	16 09 11.123	&	$-$32 06 01.20	&	7.45	&	09 Aug 2008	&	5	\\
HD~136014	&	G6~III—-IV	&	15 19 31.720	&	$-$37 05 49.78	&	6.189	&	09 Aug 2008	&	4	
\enddata
\end{deluxetable}

\clearpage

\begin{deluxetable}{cccccccccc}
\tabletypesize{\scriptsize}
\tablewidth{0pt}
\tablecolumns{10}
\tablecaption{Mean background fractions b$_{\lambda}$/s$_{\lambda}$ for Du Pont echelle spectrograph.\label{ratios_t}}
\tablehead{
\colhead{} &
\colhead{} &
\colhead{HD~142629} &
\colhead{} &
\colhead{HD~135153} &
\colhead{} &
\colhead{HD~144880} &
\colhead{} &
\colhead{HD~136014} &
\colhead{} \\
\colhead{Order} &
\colhead{$\lambda_{c}$\tablenotemark{a}} &
\colhead{Mean} &
\colhead{Mean$\times5/3$} &
\colhead{Mean} &
\colhead{Mean$\times5/3$} &
\colhead{Mean} &
\colhead{Mean$\times5/3$} &
\colhead{Mean} &
\colhead{Mean$\times5/3$} 
}
\startdata
46	&	7575  &	0.068	&	0.113	&	0.069	&	0.115	&	0.070	&	0.117	&	0.073	&	0.122	\\
47	&	7415  &  0.064	&	0.107	&	0.064	&	0.107	&	0.066	&	0.110	&	0.066	&	0.110	\\
48	&	7260  &  0.061	&	0.102	&	0.061	&	0.102	&	0.064	&	0.107	&	0.061	&	0.102	\\
49	&	7108  &  0.057	&	0.095	&	0.058	&	0.097	&	0.059	&	0.098	&	0.055	&	0.092	\\
50	&	6962  &  0.054	&	0.090	&	0.055	&	0.092	&	0.056	&	0.093	&	0.053	&	0.088	\\
51	&	6825  &  0.052	&	0.087	&	0.051	&	0.085	&	0.054	&	0.090	&	0.048	&	0.081	\\
52	&	6688  &  0.047	&	0.078	&	0.048	&	0.080	&	0.052	&	0.087	&	0.046	&	0.077	\\
53	&	6560  &  0.047	&	0.078	&	0.047	&	0.078	&	0.051	&	0.085	&	0.044	&	0.073	\\
54	&	6435  &  0.045	&	0.075	&	0.045	&	0.075	&	0.050	&	0.083	&	0.043	&	0.072	\\
55	&	6315  &  0.043	&	0.072	&	0.045	&	0.075	&	0.049	&	0.082	&	0.043	&	0.071	\\
56	&	6202  &  0.043	&	0.072	&	0.044	&	0.073	&	0.048	&	0.080	&	0.042	&	0.070	\\
57	&	6092  &  0.043	&	0.072	&	0.043	&	0.072	&	0.048	&	0.080	&	0.041	&	0.069	\\
58	&	5987  &  0.043	&	0.072	&	0.043	&	0.072	&	0.047	&	0.078	&	0.042	&	0.069	\\
59	&	5880  &  0.044	&	0.073	&	0.044	&	0.073	&	0.047	&	0.078	&	0.042	&	0.070	\\
60	&	5780  &  0.045	&	0.075	&	0.044	&	0.073	&	0.047	&	0.078	&	0.042	&	0.071	\\
61	&	5686  &  0.045	&	0.075	&	0.045	&	0.075	&	0.046	&	0.077	&	0.042	&	0.071	\\
62	&	5592  &  0.046	&	0.077	&	0.045	&	0.075	&	0.046	&	0.077	&	0.043	&	0.071	\\
63	&	5502  &  0.047	&	0.078	&	0.046	&	0.077	&	0.046	&	0.077	&	0.044	&	0.073	\\
64	&	5413  &  0.049	&	0.082	&	0.046	&	0.077	&	0.047	&	0.078	&	0.044	&	0.073	\\
65	&	5330  &  0.050	&	0.083	&	0.047	&	0.078	&	0.049	&	0.082	&	0.045	&	0.074	\\
66	&	5250  &  0.051	&	0.085	&	0.049	&	0.082	&	0.050	&	0.083	&	0.045	&	0.076	\\
67	&	5170  &  0.052	&	0.087	&	0.049	&	0.082	&	0.050	&	0.083	&	0.047	&	0.078	\\
68	&	5090  &  0.053	&	0.088	&	0.050	&	0.083	&	0.051	&	0.085	&	0.048	&	0.080	\\
69	&	5017  &  0.052	&	0.087	&	0.051	&	0.085	&	0.053	&	0.088	&	0.047	&	0.079	\\
70	&	4945  &  0.052	&	0.087	&	0.051	&	0.085	&	0.055	&	0.092	&	0.048	&	0.080	\\
71	&	4870  &  0.057	&	0.095	&	0.052	&	0.087	&	0.056	&	0.093	&	0.049	&	0.082	\\
72	&	4805  &  0.053	&	0.088	&	0.052	&	0.087	&	0.062	&	0.103	&	0.049	&	0.082	\\
73	&	4740  &  0.050	&	0.083	&	0.051	&	0.085	&	0.059	&	0.098	&	0.050	&	0.084	\\
74	&	4672  &  0.053	&	0.088	&	0.051	&	0.085	&	0.057	&	0.095	&	0.054	&	0.090	\\
75	&	4610  &  0.052	&	0.087	&	0.049	&	0.082	&	0.064	&	0.107	&	0.053	&	0.089	\\
76	&	4548  &  0.053	&	0.088	&	0.049	&	0.082	&	0.063	&	0.105	&	0.054	&	0.090	\\
77	&	4490  &  0.050	&	0.083	&	0.048	&	0.080	&	0.061	&	0.102	&	0.056	&	0.093	\\
78	&	4430  &  0.049	&	0.082	&	0.048	&	0.080	&	0.060	&	0.100	&	0.060	&	0.100	\\
79	&	4375  &  0.057	&	0.095	&	0.050	&	0.083	&	0.063	&	0.105	&	0.045	&	0.074	\\
80	&	4320  &  0.050	&	0.083	&	0.047	&	0.078	&	0.055	&	0.092	&	0.054	&	0.090	\\
81	&	4265  &  0.044	&	0.073	&	0.044	&	0.073	&	0.057	&	0.095	&	0.053	&	0.088	\\
82	&	4210  &  0.046	&	0.077	&	0.045	&	0.075	&	0.064	&	0.107	&	0.047	&	0.078	\\
83	&	4160  &  0.047	&	0.078	&	0.048	&	0.080	&	0.069	&	0.115	&	0.058	&	0.097	\\
84	&	4110  &  0.053	&	0.088	&	0.046	&	0.077	&	0.067	&	0.112	&	0.063	&	0.105	\\
85	&	4060  &  0.046	&	0.077	&	0.046	&	0.077	&	0.070	&	0.117	&	0.062	&	0.104	\\
86	&	4012  &  0.049	&	0.082	&	0.044	&	0.073	&	0.082	&	0.137	&	0.099	&	0.166	\\
87	&	3966  &  0.057	&	0.095	&	0.047	&	0.078	&	0.123	&	0.205	&	0.099	&	0.165	\\
88	&	3920  &  0.061	&	0.102	&	0.046	&	0.077	&	0.100	&	0.167	&	0.105	&	0.175	
\enddata
\tablenotetext{a}{Central wavelength of the order.}
\end{deluxetable}

\clearpage

\begin{deluxetable}{lcccc}
\tabletypesize{\scriptsize}
\tablewidth{0pt}
\tablecolumns{5}
\tablecaption{Radial Velocities\label{RVs}}
\tablehead{
\colhead{Star} &
\colhead{HJD at midpoint} & 
\colhead{Phase}  & 
\colhead{RV}  & 
\colhead{err}  \\
\colhead{} &
\colhead{(2450000+)} &
\colhead{($\phi$)} & 
\colhead{(km s$^{-1}$)}  & 
\colhead{(km s$^{-1}$)}  
}
\startdata
CD Vel	&	3836.48565	&	0.00	&	210.76	&	0.50	\\
	&	3836.49453	&	0.02	&	210.97	&	0.48	\\
	&	3836.54295	&	0.10	&	216.80	&	0.35	\\
	&	3836.54928	&	0.11	&	218.06	&	0.35	\\
               &       .       &       .       &       .       &       .     \\
               &       .       &       .       &       .       &       .     \\
               &       .       &       .       &       .       &       .      
\enddata
\tablecomments{Table~\ref{RVs} is published in its entirety in the
electronic edition of the {\it Astrophysical Journal Supplement Series}.  
A portion is shown here for guidance regarding its form and content.}
\end{deluxetable}

\clearpage

\begin{deluxetable}{lcccccc}
\tabletypesize{\scriptsize}
\tablewidth{0pt}
\tablecolumns{7}
\tablecaption{Ephemerides of our program stars\label{eph}}
\tablehead{
\colhead{Star} &
\colhead{Data used} & 
\colhead{$T_{0}$\tablenotemark{a}}  & 
\colhead{err}  & 
\colhead{Period\tablenotemark{b}} &
\colhead{Period} &
\colhead{error}  \\
\colhead{} &
\colhead{(HJD 2450000+)} &
\colhead{(HJD 2450000+)} & 
\colhead{(HJD 2450000+)} & 
\colhead{(day)}  &
\colhead{(day)}  &
\colhead{(day)}  
}
\startdata
CD Vel	&	all	&	3837.632	&	0.0003	& 0.57351	& 0.573510	& 0.000003\\
WY Ant	&	all	&	4191.685	&	0.0097	& 0.57434	& 0.574344	& 0.000002\\
DT Hya	&	all	&	4583.637	&	0.0089	& 0.56797	& 0.567978	& 0.000001\\
AS Vir	&	all	&	4907.709	&	0.0098	& 0.553439	& 0.553412	& 0.000002\\
RV Oct	&	all	&	3841.602	&	0.0016	& 0.571184	& 0.571170	& 0.000002\\
XZ Aps	&	all	&	3842.735	&	0.0052	& 0.5873	& 0.587264	& 0.000002\\
BS Aps	&	all	&	4583.785	&	0.0045	& 0.582577	& 0.582561	& 0.000007\\
UV Oct	&	3836.84—3842.91, 4306.46—5021.84	&	3837.875	&	0.0072	& 0.542561 &	0.542578 & 0.000003	\\
	&	3931.58—4194.92, 5070.48—5073.59	&	5070.605	&	0.0072	& \nodata  &    \nodata	 & \nodata	\\
V1645 Sgr	&	4191.89—4306.90	&	4306.775	&	0.0150	& 0.552979 &        0.552948 & 0.000005	\\
	&	4579.85—4583.91	&	4579.895	&	0.0150	&	\nodata    &  \nodata	& \nodata \\
	&	3932.73—3946.75, 4687.66—5074.71	&	4687.703	&	0.0170	& \nodata & \nodata & \nodata		\\
Z Mic	&	all	&	5075.606	&	0.0015	& 0.58693 &	0.586926 & 0.000001	\\
TY Gru	&	3933.79—3935.65, 5071.50--5073.66	&	3933.785	&	0.0120	& \nodata &	0.570065 & 0.000005	\\
	&	3945.63—4306.89	&	4304.885	&	0.0120	&	\nodata & \nodata & \nodata	
\enddata
\tablenotetext{a}{Epoch at time of light maxima or radial velocity minima.}
\tablenotetext{b}{Listed in ASAS catalog.}
\end{deluxetable}


\end{document}